\documentclass[trackchanges, twocolumn,twocolappendix]{aastex701}
\usepackage{amsmath}

\begin{document}

\title{First Principles Magnetohydrodynamical Theory for the Expanding Box Model: Effects on Expansion-Induced Alfvén Wave Reflection}

\author[orcid=0009-0000-5670-0522]{Sebastián Saldivia}
\affiliation{Departamento de Física, Facultad de Ciencias, Universidad de Chile, Las Palmeras 3425, Santiago, 7800003, Chile}
\email[show]{sebastian.saldivia@ug.uchile.cl}  

\author[0000-0003-4887-9512]{Nicol\'as Villarroel-Sep\'ulveda}
\affiliation{Departamento de F\'isica, Facultad de Ciencias F\'isicas y Matem\'aticas, Universidad de Concepci\'on, Concepci\'on 4070386, Chile}
\email{}  

\author[orcid=0000-0001-8103-017X]{Sebastián Echeverría-Veas}
\affiliation{Departamento de Física, Universidad de Santiago de Chile, Santiago 9170020, Chile}
\email{sebastian.echeverria@ug.uchile.cl}

\author[0000-0002-7085-658X]{Felipe A. Asenjo}
\affiliation{Facultad de Ingeniería y Ciencias, Universidad Adolfo Ibáñez, Santiago 7491169, Chile}
\email{felipe.asenjo@uai.cl}

\author[0000-0002-9161-0888]{Pablo S. Moya}
\affiliation{Departamento de Física, Facultad de Ciencias, Universidad de Chile, Las Palmeras 3425, Santiago, 7800003, Chile}
\email[show]{\\pablo.moya@uchile.cl}

\begin{abstract}

The Expanding Box Model (EBM) has been widely employed to simulate multiscale plasma phenomena in the expanding solar wind by transforming the MHD equations to a co-moving, non-inertial frame. However, traditional formulations have suffered from historical ambiguity regarding the physical separation between the co-moving and inertial reference frames, primarily arising from a classical approximation of an invariant magnetic field between them. To resolve this inconsistency, we reformulate the EBM from first principles using a fully covariant approach. Here, we model the expanding solar wind frame as an anisotropic expanding spacetime metric, allowing us to incorporate radial acceleration profiles and differential transverse expansion, ensuring that all physical fields are correctly transformed by expansion. We demonstrate that asymmetries identified in previous EBM-MHD literature are direct consequences of neglecting the tensorial scaling of the magnetic field. Our covariant treatment eliminates these residues, restoring symmetry in the co-moving frame. Projecting our system back into the inertial frame clarifies the distinction between local plasma dynamics and plasma expansion, revealing the anisotropy of the Parker spiral as a geometric projection. Furthermore, linear wave analysis using Els\"asser variables reveals that plasma expansion induces a low-frequency cutoff and geometric damping. Numerical integration demonstrates that expansion drives the reflection of Alfvén waves, generating counter-propagating modes primarily at low frequencies relative to the expansion rate, while high frequencies converge to the WKB approximation. This provides a consistent foundation for simulations, establishing that expansion can serve as a source of counter-propagating waves necessary to drive solar wind turbulence at low frequencies.

\end{abstract}



\section{Introduction}
\label{sec1}

One of the great current problems in heliophysics is understanding how the solar wind is heated as it expands~\citep{Bruno2013, Verscharen2019}. As the plasma drifts away from the Sun, it is subject to several processes, ranging from turbulent heating and wave-particle interactions at microscopic scales to expansion-induced cooling at macroscopic scales. Exactly how these multiscale processes interact to shape the plasma properties observed from the inner heliosphere to near-Earth orbit remains a central challenge in modern space physics~\citep{Raouafi2023,Rivera2024}. In this context, the Expanding Box Model (EBM) was proposed by ~\citet{Velli1992} and ~\citet{Grappin1993} to tackle this problem by incorporating the effects of radial plasma expansion into the magnetohydrodynamic (MHD) equations. This is achieved through a coordinate transformation to a non-inertial reference frame that co-moves with the solar wind parcel. In this frame, the spherical expansion is locally approximated using a Cartesian coordinate system, in which the radial coordinate is transformed via a Galilean boost and the transverse coordinates are renormalized, thereby ensuring a plasma box with constant volume. Thus, in the EBM framework, plasma expansion appears as additional non-inertial forces that modify the MHD equations, and the spatial derivatives in the co-moving frame account for plasma stretching ~\citep{Grappin1993, Grappin1996}. This formulation is particularly useful for numerical simulations, reducing memory constraints for expanding plasma phenomena, as the effects of expansion are expressed through additional non-inertial forces rather than by explicitly growing simulation parcels.

Over the past few decades, the EBM has become increasingly relevant in heliospheric physics, motivating extensive research into the role of plasma expansion in solar wind heating dynamics over kinetic and fluid scales~\citep{Matteini2006,Ofman2011,Moya2012,Micera2021}. It has been utilized in hybrid simulations to study Alfvén wave propagation ~\citep{Liewer2001}, Alfvénic fluctuations~\citep{Matteini2024}, and proton firehose instabilities~\citep{Hellinger2008,Hellinger2017}. The EBM has been employed to study the radial evolution of MHD turbulence~\citep{Dong2014,Verdini2024,Shi2025}, magnetic clouds~\citep{Sangalli2025}, and the parametric decay of Alfvén waves~\citep{Tenerani2013,Del_Zanna2015}. The model has also been utilized to perform fully kinetic simulations~\citep{Innocenti_2019,Innocenti2020} and to study Alfvénic fluctuations in a fast-slow stream interaction in the solar wind~\citep{Shi2020}. The EBM has also proven to be a powerful tool in theoretical work, since including expansion as a non-inertial force in the MHD equations offers several analytical advantages. In this context, the EBM has been employed to study nonlinear Alfvén wave propagation~\citep{Nariyuki2015}, the double-adiabatic evolution of solar wind~\citep{Echeverria-Veas2024}, and the dispersion properties of expanding MHD normal modes~\citep{Saldivia2025}. Furthermore, the model has been generalized to an Accelerating Expanding Box (AEB) to include radial solar wind acceleration near the Sun~\citep{Tenerani2013,Tenerani2017}, to a quasi-linear EBM to characterize microscopical scales~\citep{Seough2023}, and to account for reduced magnetohydrodynamic turbulence~\citep{David2025}. 

In this context, the EBM has become a valuable tool for studying expanding plasma phenomena, enabling the modeling of multiscale solar wind dynamics with great success over the last few decades. However, standard EBM formulations have historically relied on a Galilean coordinate transformation in which the magnetic field is Galilean-invariant (the magnetic limit in Galilean electromagnetism)~\citep{Velli1992, Grappin1993, Liewer2001}. Nevertheless, this approximation breaks if a coordinate's renormalization is coupled to the Galilean boost. This approximation has persisted even in deeper kinetic derivations, where the fluid EBM equations are obtained by taking the fluid moments of an expanding Vlasov Equation~\citep{Echeverria-Veas2023}. Because the fields are transformed assuming only a Galilean boost, the macroscopic expansion of the coordinates structurally alters the spatial differential operators. This subtle approximation has created a conceptual ambiguity in the literature, as the physical distinction between an inertial frame, in which actual spacecraft measurements are recorded, and the non-inertial co-moving frame, in which simulations are performed, has not been systematically defined previously. Specifically, expansion terms remained explicit in the spatial differential operators for magnetic tension and magnetic pressure, as seen in ~\citet{Echeverria-Veas2023}. For numerical implementations, these artifacts can complicate the evaluation of conservative fluxes, breaking the symmetry of ideal MHD. As a consequence, there is a fundamental need to define the differences between physical observables across different reference frames. Clarifying the physical distinction between the non-inertial co-moving frame and the inertial laboratory frame can provide the theoretical foundation required to directly apply these expanding models to actual \textit{in situ} spacecraft measurements.  

To resolve this historical ambiguity, solar wind expansion must be modeled within a framework that embeds physical vector fields in an expanding coordinate system, ensuring that vector norms scale correctly with the expanding geometry. Performing a rigorous transformation of the physical vector fields allows for a clear distinction between the physics of the non-inertial and inertial frames, resolving ambiguities in the literature and providing a set of equations in each frame that accurately represent the expanding plasma dynamics. 
To achieve this purpose, we formulate the EBM by describing the non-inertial forces using an expanding spacetime metric, in analogy to an anisotropic cosmological model. The interpretation of EBM in terms of an evolving metric using a general relativistic approach was first pioneered by \citet{DelZanna2012}. Their work, together with subsequent implementations \citep{Del_Zanna2015, Toci2018}, established the foundation for modeling solar wind expansion in a covariant framework. This work presents a generalization of this approach, accounting for both the proper transformation of fields, as pioneered by \citet{DelZanna2012}, and the inclusion of radial solar wind acceleration. The radial acceleration has been incorporated in previous non-GR EBM frameworks \citep{Tenerani2013,Nariyuki2015,Tenerani2017}, but here we incorporate it directly into the metric tensor to perform the EBM transformations consistently. While the solar wind speed is below relativistic speeds, we employ the covariant approach because it provides a robust mathematical formalism to describe an expanding spacetime in which vector fields can be projected between an inertial space and a stretching domain, and also generalizes to other expanding astrophysical phenomena.  On the other hand, given the proliferation of different sets of EBM-MHD equations published over the last few decades, there is a fundamental need to clarify which formulations retrieve a consistent set of equations. 

This article is organized as follows: in Section \ref{sec2}, we introduce the formalism for describing radial plasma expansion while ensuring the correct transformation of vector fields between reference frames. In Section \ref{sec3}, we present the system of expanding MHD equations deduced from this framework. We present this system of equations in both the co-moving and non-inertial frames, and study the physical equivalence between the two, including linear analysis of Alfvén waves as solutions of the inertial system. In Section \ref{sec4}, we present the Els\"asser formulation of the ideal, expanding MHD system of equations. In Section \ref{sec5}, we present numerical integrations to validate our model over different wave frequency regimes. Finally, in Section \ref{sec6}, we summarize our results and present conclusions.

\section{Covariant Formalism}
\label{sec2}

To systematically resolve the frame ambiguities previously discussed, we model the accelerating, expanding solar wind domain as an anisotropic Bianchi Type I spacetime metric~\citep{MTW}, in which expansion is a geometric property of space itself. This allows the coordinate space itself to expand with the plasma, in analogy to MHD in an expanding universe~\citep{Jacobs1969,Dunn1980}. By encoding the coordinate transformations directly into the spacetime metric, we systematically deduce the ideal EBM-MHD equations. By evaluating the covariant conservation of mass, the energy-momentum tensor, and Maxwell's equations following standard MHD formalisms~\citep{Lichnerowicz1967, Anile1990}, we take the non-relativistic limit appropriate for a solar wind-like environment. This approach ensures that all vector fields and their governing differential operators remain physically and mathematically consistent with the stretching plasma coordinate transformations. The full covariant derivation is detailed in Appendix \ref{appendixA}, and this section establishes the geometric scaling of physical fields. 

We consider a plasma parcel moving radially away from the Sun. Following a Cartesian approximation, we define an inertial, static laboratory frame $S$ with coordinates $x^{\mu} = (t,x,y,z)$ centered at the Sun. Following the standard EBM formulation~\citep{Velli1992,Grappin1996}, we introduce a non-inertial frame $S'$ with coordinates $x^{\mu'} = (t',x',y',z')$ that co-moves with the plasma parcel. The plasma has a radial velocity $V(t')$ and its heliocentric distance is $R(t')$. These profiles can be kept general, allowing the framework to implement uniformly accelerated profiles~\citep{Tenerani2013,Nariyuki2015} or to include a jerk in the velocity to better match solar wind observations~\citep{Halekas2022}.
While our covariant framework naturally accommodates relativistic kinematics, the solar wind reaches a terminal velocity far below the speed of light. Thus, we consider a non-relativistic expansion of the solar wind $V(t') \ll c$ before evaluating the dynamics.

To maintain a constant volume in the co-moving frame $S'$, the physical stretching of the plasma must be absorbed by the coordinate geometry.
For an accelerating solar wind, the stretching of a fluid parcel in the radial direction is proportional to its bulk speed, while its transverse expansion rate is proportional to the heliocentric distance. We represent this stretching through two scale factors, the longitudinal expansion rate $w(t') = V(t')/V_0$ and the transverse expansion rate $a(t') = R(t')/R_0$, where $V_0$ and $R_0$ are the initial velocity and position, respectively. As the physical distances within the plasma box are stretched as $dx = w(t') dx'$, $dy = a(t') dy'$ and $dz = a(t') dz'$, in the co-moving frame $S'$, the expanding geometry dictates that the spatial basis vectors $\mathbf e_{i'}$ of the co-moving frame are not unitary, as their physical lengths stretch proportionally to the anisotropic expansion ($||\mathbf e_{x'}|| = w(t')$, $||\mathbf e_{y'}|| = ||\mathbf e_{z'}|| = a(t')$).
\begin{figure}[ht!]
    \centering
    \includegraphics[width=0.9\linewidth]{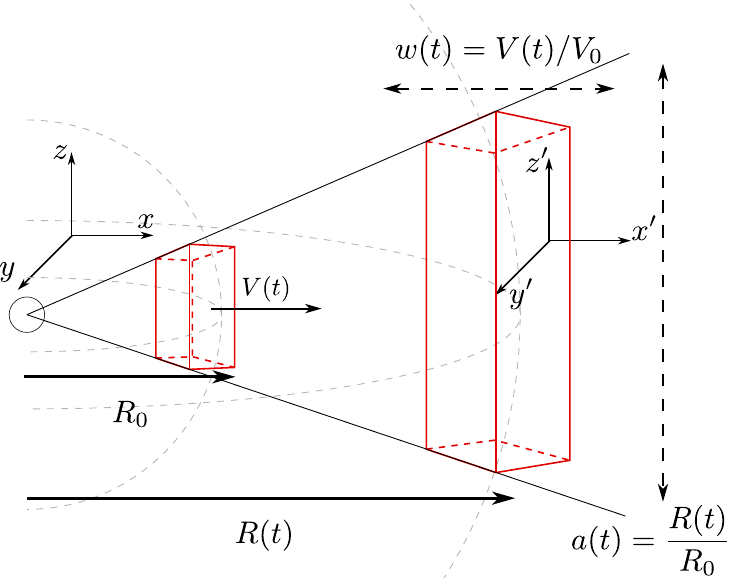}
    \caption{Schematic representation of the EBM geometry in the Cartesian approximation. The inertial laboratory frame $S$ observes a radially expanding plasma moving away at $R(t')$, while the non-inertial, co-moving frame $S'$ maintains a constant box volume through the renormalization by the longitudinal $w(t')$ and transverse $a(t')$ scale factors.}
    \label{ebm}
\end{figure}
This geometry governs the transformation of plasma quantities between reference frames.
As the co-moving basis vectors $\mathbf{e}_{i'}$ are stretched with the expansion, the contravariant components of physical vectors, such as the plasma velocity, must shrink proportionally in the co-moving frame to represent the same physical quantity. 
This scaling is deduced from the stretching of basis vectors, as shown in Appendix \ref{appendixa1}, Equation \eqref{A5}. On the other hand, as the magnetic field $\mathbf{B}$ is a pseudovector, its transformation is derived from the spatial projection of the electromagnetic tensor $F_{\mu\nu}$, as detailed in Appendix \ref{a2}.

Consequently, the plasma bulk velocity, the magnetic field, and the electric field transform between frames as
\begin{align}
    \mathbf{v} = \mathbb{A} \cdot \mathbf{v}', && \mathbf{B} = \mathbb{A} \cdot \mathbf{B}', && \mathbf{E} = \mathbb{A} \cdot \mathbf{E}', \label{vector_transformations}
\end{align}
where
\begin{equation}
    \mathbb A = \begin{pmatrix}
        w(t')& 0 &0\\
        0 & a(t') & 0\\
        0 & 0 & a(t')
    \end{pmatrix}.
\end{equation}
Here, the transformation for the plasma bulk velocity coincides with \citet{Echeverria-Veas2023}. However, our covariant transformation of the electric and magnetic fields marks a departure from the EBM treatments in the literature. In previous works, the electromagnetic fields are typically transformed via a Galilean boost, either strictly to the electric field~\citep{Echeverria-Veas2023} or to both fields~\citep{Innocenti_2019}. Nevertheless, these frameworks treat the magnetic field as invariant between frames ($\mathbf{B} = \mathbf{B}'$). By neglecting the stretching of the basis vectors, previous frameworks modeled a magnetic field that does not explicitly consider EBM coordinate transformations. In our covariant approach, every vector field is naturally coupled to the expansion. As the space expands, the vector components change in response to the expansion. On the other hand, for any scalar quantity $\phi$ (a rank-0 tensor, such as the proper mass density $\rho$ or pressure $p$), the transformation is trivial, as $\phi = \phi'$. 

Finally, the spatial differential operator $\nabla'$ transforms according to the chain rule of partial derivatives ($\partial_{i'} = \frac{\partial x^i}{\partial x^{i'}} \partial_i$). Since the inertial coordinates scale as $x = w x'$, $y = a y'$ and $z = a z'$, the spatial gradients transform inversely to the basis vectors, as
\begin{equation}
    \nabla = \mathbb{A}^{-1} \cdot \nabla'.
\end{equation}
This transformation is consistent with the literature ~\citep{Grappin1993, Nariyuki2015,Echeverria-Veas2023}.
Applying these explicit transformations to the covariant conservation laws, specifically mass continuity, energy-momentum conservation, and Faraday's Law, yields the full ideal EBM-MHD system from first principles in the expanding spacetime. The full derivation of the relativistic EBM-MHD equations is detailed in Appendix \ref{a2}. In the following section, we formulate the non-relativistic equations governing the expanding solar wind directly from this geometric foundation.

\section{EBM-MHD Equations}
\label{sec3}

Having established the correct geometric frame transformations in Section \ref{sec2} that allow us to deduce the relativistic, EBM-MHD set of equations directly from the covariant conservation of mass, the energy-momentum tensor, and Maxwell's equations (as detailed in Appendix \ref{a2}), in this section we take the non-relativistic limit of Equations \eqref{eq_cont_rel}-\eqref{eq_press_rel} to obtain the set of ideal EBM-MHD equations from a covariant approach. This first-principles foundation guarantees that our resulting EBM-MHD system is free from geometric residues that emerge when expanding spatial operators are mixed with rigid vector fields, making a clear physical distinction between the non-inertial and inertial frames.

Here, we first present the non-relativistic EBM-MHD equations evaluated within the co-moving frame $S'$. In this non-inertial system, the computational volume remains constant, and the solar wind expansion is encoded in the source terms, restoring the topological symmetry of the ideal MHD operators for numerical implementations. Subsequently, we project these equations back into the inertial frame $S$ to demonstrate their structural equivalence with classical literature and physical observables.

\subsection{Co-moving Frame}

The non-relativistic limit of the EBM-MHD equations is achieved by assuming that the plasma bulk velocity is much smaller than the speed of light ($v \ll 1$), causing all relativistic effects to vanish. Additionally, assuming the thermal and magnetic energy densities are negligible compared to the rest-mass energy density ($p \ll \rho c^2$, $B^2 \ll \rho c^2$), the system of equations simplifies significantly. Applying these limits to the covariant equations \eqref{eq_cont_rel}-\eqref{eq_press_rel}, we obtain the non-relativistic EBM-MHD equations in the co-moving frame $S'$. The full set of ideal MHD equations reduces to
\begin{equation}
    \frac{\partial \rho}{\partial t'} + \nabla'\cdot (\rho \mathbf v') = - \text{Tr}(\mathbb H) \rho,\label{eq_cont_nr}
\end{equation}
\begin{equation}
    \begin{split}
         \frac{\partial \mathbf B'}{\partial t'} +  (\nabla' \cdot \mathbf v')\mathbf B'  + (\mathbf v' \cdot  \nabla') \mathbf B'\\ - (\mathbf B' \cdot \nabla')\mathbf v' = - \text{Tr}(\mathbb H) \mathbf  B'\label{eq_faraday_nr}
    \end{split}
\end{equation}
\begin{equation}
\begin{aligned}[b]
    \rho \frac{\partial \mathbf{v}'}{\partial t'} + \rho (\mathbf{v}' \cdot \nabla')\mathbf{v}' + g^{i'j'} \nabla' p + \frac{1}{8 \pi} g^{i'j'} \nabla' B^2 \\- \frac{1}{4 \pi}(\mathbf{B}' \cdot \nabla')\mathbf{B}' = - 2 \rho \mathbb{H} \cdot \mathbf{v}'.  
\end{aligned}
\label{eq_mom_nr}
\end{equation}
\begin{equation}
    \frac{\partial p}{\partial t'} + (\mathbf v' \cdot \nabla') p + \Gamma p(\nabla'\cdot \mathbf v') = - \Gamma p \text{Tr}(\mathbb H),\label{eq_press_nr}
\end{equation}
where $\Gamma$ is the plasma polytropic index, $g^{i'j'}$ is the expanding metric defined in Equation \eqref{metric} and  $  \mathbb H = \text{diag}\left (\frac{\dot w}{w}, \frac{ \dot a}{a}, \frac{\dot a}{a} \right )$
is the expansion rate tensor, in analogy to \citet{Nariyuki2015}. Defining the expansion and acceleration rates $\gamma_a = \dot a/a$ and $\gamma_w = \dot w/w$, the expansion rate matrix trace is $\text{Tr}(\mathbb H) = 2 \gamma_a + \gamma_w$. 

Equations \eqref{eq_cont_nr}-\eqref{eq_press_nr} correspond to the continuity equation, Faraday's law, momentum equation, and pressure equation, respectively, for an expanding, non-relativistic plasma. Note that these equations are written in the non-inertial frame and in vector form for easier comparison with the literature. 

This covariant deduction reveals a structural advantage compared with previous EBM formulations. By properly accounting for the geometric transformation of the vector fields, the expansion matrices associated with the stretching basis exactly cancel the inverse matrices originating from the spatial gradients. As a consequence, the left-hand side of the equations retains the exact shape of standard, ideal MHD. This ensures that terms such as magnetic tension remain equivalent to those in the non-expanding case. Consequently, all the information about expansion is expressed in the non-inertial terms on the right-hand side. This is in contrast to previous formulations~\citep{Echeverria-Veas2023}, where explicit $\mathbb A^{-1}$ terms appeared in the magnetic field and momentum equations, as a consequence of maintaining the magnetic field as invariant to the expansion. 

Furthermore, our formalism explicitly captures how the expanding metric modifies the isotropic forces. In the momentum equation \eqref{eq_mom_nr}, the gradients of the thermal and magnetic pressures ($\nabla' p$ and $\nabla' B^2$, respectively) are contracted by the metric tensor $g^{i'j'}$. Because these total-pressure terms are scalar invariants, their magnitudes do not scale with the plasma. Thus, transforming their local gradients requires raising the spatial index via the metric. This process introduces geometric weights ($w^{-2}$ and $a^{-2}$), showing how the expanding spacetime anisotropically dilutes the effective pressure forces acting on the plasma. Similar to the standard formulation, the effect of plasma expansion is represented in the equations by non-inertial forces on the right-hand side. 

Finally, our co-moving momentum equation \eqref{eq_mom_nr} introduces a factor of 2 in this geometric drag term ($- 2 \rho \mathbb{H} \cdot \mathbf{v}'$) that is absent in the literature. This term captures the full inertial effect of the accelerating expansion on the momentum flux, providing a geometric origin for the Accelerating Expanding Box (AEB) regime \citep{Tenerani2013}. This complete non-inertial coupling is partially lost when the expanding equations are derived strictly from kinematic coordinate transformations on flat-space MHD.

\subsection{Inertial Frame}

We can assess the consistency between our covariant deduction and the literature by transforming Equations \eqref{eq_cont_nr}-\eqref{eq_press_nr} to the inertial frame $S$. This coordinate system represents the physical space where an observer measures the solar wind expanding away from a fixed point. Mapping our co-moving equations back to this measurable domain requires applying the inverse spatial transformation $\nabla = \mathbb A^{-1} \cdot \nabla'$ and $\mathbf B = \mathbb A \cdot \mathbf B'$, $\mathbf v = \mathbb A \cdot \mathbf v'$, as discussed in Section \ref{sec2}.

By applying these transformations to all vector fields and operators (as detailed in Appendix \ref{appendixA}), we obtain the EBM-MHD equations in the inertial frame $S$:
\begin{equation}
    \frac{\partial \rho}{\partial t} + \nabla \cdot (\rho \mathbf{v}) = -\text{Tr}(\mathbb{H})\rho, \label{eq_cont_s} 
\end{equation}
\begin{equation}
    \begin{aligned}[b]
        \frac{\partial \mathbf{B}}{\partial t} + (\nabla \cdot \mathbf{v})\mathbf{B} + (\mathbf{v} \cdot \nabla) \mathbf{B} \\- (\mathbf{B} \cdot \nabla) \mathbf{v} = - \left [\frac{\dot a}{a} \mathbb{L} + \frac{\dot w}{w} \mathbb{T}\right ]\cdot \mathbf{B}, \label{eq_faraday_s}
    \end{aligned}
\end{equation}
\begin{equation}
    \begin{aligned}[b]
        \rho \frac{\partial \mathbf{v}}{\partial t} + \rho (\mathbf{v} \cdot \nabla) \mathbf{v} + \nabla p + \frac{1}{8 \pi}\nabla B^2\\ - \frac{1}{4 \pi}(\mathbf{B} \cdot \nabla) \mathbf{B} = - \rho \mathbb{H} \cdot \mathbf{v}, \label{eq_mom_s} 
    \end{aligned}
\end{equation}
\begin{equation}
\frac{\partial p}{\partial t} + (\mathbf{v} \cdot \nabla) p + \Gamma p(\nabla\cdot \mathbf{v}) = - \Gamma p \text{Tr}(\mathbb{H}), \label{eq_press_s} 
\end{equation}
where $\mathbb{L} = \text{diag}(2,1,1)$ and $\mathbb{T} = \text{diag}(0,1,1)$. 

The full set of Equations \eqref{eq_cont_s}-\eqref{eq_press_s} reveals the EBM-MHD equations in the inertial frame as deduced from a covariant approach. Our results for the continuity \eqref{eq_cont_s} and pressure \eqref{eq_press_s} equations are equivalent to those found in previous works~\citep{Grappin1996, Nariyuki2015,Echeverria-Veas2023}. Since scalar fields are rank-0 tensors, they do not couple to the expanding basis vectors. Thus, previous works based on coordinate transformations captured the correct macroscopic evolution for the pressure and density.

The significance of our derivation is clearly reflected in the vector equations. Our inertial Faraday's law is equivalent to the one deduced by \citet{Tenerani2013} and \citet{Nariyuki2015}. Our covariant formulation offers an equivalent, complementary pathway, where the same terms arise inherently from the Christoffel connections of the expanding metric. This equivalence shows that macroscopic flux conservation is encoded within the geometric structure of the expanding spacetime.

Furthermore, this proper geometric mapping reveals the origin of the structural anomalies found in the differential operators of previous co-moving models. In classical EBM derivations, authors transformed the spatial operators from the inertial to the co-moving frame while treating the magnetic field as invariant between frames. As a consequence, works such as \citet{Echeverria-Veas2023} obtained a co-moving Faraday's law where the inverse expansion matrix remained structurally trapped within the spatial derivatives, evaluating to $[\mathbf B' \cdot (\mathbb A^{-1 }\cdot \nabla')] (\mathbb A \cdot \mathbf v') $, with similar terms in the momentum equation. Identical mathematical artifacts implicitly appear in the magnetic tension operators of \citet{Grappin1996} and subsequent literature. In our model, the tensor scaling of the physical fields compensates for the deformation of the spatial gradients. Through this transformation, our framework eliminates these historical artifacts.

\subsection{Physical Equivalence }

Despite showing that the magnetic tension and pressure terms in previous EBM formulations contain geometric artifacts as a result of not correctly transforming the magnetic field between frames, the physical observables ($\mathbf v$, $\mathbf B$) measured in the inertial frame $S$ remain identical in both frameworks.  

In the heliosphere, it is a well-established observational fact that the interplanetary magnetic field follows an anisotropic Parker spiral structure \citep{Parker1958}. For a solar wind expanding at a constant radial velocity ($w(t') = 1$), the radial magnetic field decays as $B_{x} \propto 1/R^2$, while the transverse components decay as $B_{y,z} \propto 1/R$. Because previous formulations relied on assuming $\mathbf{B}' = \mathbf{B}$, the simulated co-moving magnetic field in $S'$ artificially retained this exact inertial anisotropy. As previous work did not geometrically scale the vector fields before mapping the spatial operators ($\nabla = \mathbb{A}^{-1} \cdot \nabla'$), the inverse expansion matrices were maintained within the differential operators, leading to the mathematical artifacts appearing in the momentum equation and Faraday's law in the literature~\citep{Echeverria-Veas2023}.

Our covariant deduction solves this problem by identifying exactly where the anisotropy originates. By properly applying the geometric scale matrix ($\mathbf{B} = \mathbb{A} \cdot \mathbf{B}'$), Equation \eqref{eq_faraday_nr} dictates a purely isotropic local decay ($\mathbf{B}' \propto 1/a^2$) within the co-moving grid. Thus, it is the expanding metric tensor itself that transforms the isotropic local field of the $S'$ frame into the anisotropic Parker spiral observed by spacecraft in the inertial frame $S$. By identifying the purely geometric origin of this anisotropy, our model provides a robust, non-inertial coordinate system to simulate expanding plasma phenomena. It recovers the correct physical observables in the measurable space while preserving the symmetry of the ideal MHD equations within the co-moving box.

In Appendix \ref{appendixD}, we present a systematic comparison of these equations with several deductions presented in the literature.

\section{Compressible Els\"asser Variables Formulation}
\label{sec4}

To provide a theoretical foundation for future numerical simulations and the theoretical study of nonlinear interactions in the accelerating solar wind, it is highly advantageous to formulate the EBM-MHD system in terms of Els\"asser variables~\citep{Elsasser1950}. In standard MHD, transforming the system of equations using these variables yields a symmetric system that explicitly separates outward- and inward-propagating Alfvénic fluctuations. Because the turbulent cascade in plasmas is driven mainly by interactions between counter-propagating wave packets~\citep{Dobrowolny1980}, Els\"asser variables constitute the most natural framework for characterizing fully developed MHD turbulence, energy dissipation, and cross-helicity evolution~\citep{Tu1995, Bruno2013}.

Defining the magnetic field in Alfvénic units as $\mathbf b = \mathbf B/\sqrt{4 \pi \rho}$, we introduce the Els\"asser variables as
\begin{equation}
    \mathbf z^{\pm} = \mathbf v \pm \mathbf b.
\end{equation}
These variables represent Alfvénic perturbations propagating in the direction of the background magnetic field $(\mathbf z^+)$ or in the opposite direction $(\mathbf z^-)$. Following \cite{Marsch1987}, Equations \eqref{eq_cont_s}-\eqref{eq_press_s} can be written for compressible medium variables, allowing density perturbations ($\nabla \rho \neq 0$). By adding and subtracting the magnetic field condition $\nabla \cdot \mathbf B = 0$ to Equation \eqref{eq_cont_s} and expanding in Alfvénic units, allowing for density compressibility, the continuity equation can be written as
\begin{equation}
    \frac{D^\pm}{Dt} \ln \rho = - \nabla \cdot (\mathbf z^\pm \pm \mathbf b) - \text{Tr}(\mathbb H),\label{elsasser1}
\end{equation}
where we have defined the convective derivative along the Els\"asser fields as $
    \frac{D^{\pm}}{Dt} = \frac{\partial}{\partial t} + (\mathbf z^{\pm} \cdot \nabla)$.
Equation \eqref{elsasser1} describes the evolution of the plasma density in the expanding frame. Similarly, adding and subtracting Equation \eqref{eq_faraday_s} from \eqref{eq_mom_s}, we obtain the generalized compressible Els\"asser equation for the expanding, accelerating solar wind:
\begin{equation}
    \frac{D^\mp \mathbf z^\pm}{Dt} = - \frac{1}{\rho} \nabla P_T \mp \frac{1}{2} \mathbf b [\nabla \cdot (\mathbf z^\pm \pm \mathbf b)] - \mathbb H \cdot \mathbf v \mp \mathbb K \cdot \mathbf b,
\end{equation}
where $\mathbb K = -\frac{1}{2} \text{Tr} (\mathbb H) + \frac{\dot a}{a} \mathbb L + \frac{\dot w}{w} \mathbb T$. By expanding the old variables as $\mathbf v =  (\mathbf z^+ + \mathbf z^-)/2$ and $\mathbf b = (\mathbf z^+ - \mathbf z^-)/2$, the complete and self-contained ideal EBM-MHD equations in terms of Els\"asser variables takes its final, symmetric form:
\begin{equation}
    \frac{D^\pm}{Dt} \ln \rho = - \frac{1}{2}\nabla \cdot (3 \mathbf z^\pm -\mathbf z^\mp) - \text{Tr}(\mathbb H),\label{elsasser0}
\end{equation}
\begin{equation}
\begin{aligned}[b]
        \frac{D^\mp \mathbf z^\pm}{Dt} = \pm \frac{1}{4} (\mathbf z^+ - \mathbf z^-) \left (\frac{D^\pm}{Dt} \ln \rho + \text{Tr}(\mathbb H)\right ) \\-  \frac{1}{8}  \nabla (\mathbf z^+ - \mathbf z^-)^2 - \left ( \tilde v_S^2 + \frac{1}{8} (\mathbf z^+ - \mathbf z^-)^2 \right )\nabla \ln \rho\\- \frac{1}{4} \text{Tr}(\mathbb H) \mathbf z^\pm - \mathbb M \cdot \mathbf z^\mp.\label{elsasser2}
\end{aligned}
\end{equation}
Here, $\tilde{v}_S \propto (wa^2)^{-(\Gamma - 1)/2}$ represents the local sound speed, which evolves with the radial distance due to the polytropic closure in Equation \eqref{eq_press_s}.
Expansion introduces a geometric damping represented by the $-\frac{1}{4} \text{Tr}(\mathbb H) \mathbf z^\pm$ term.
Furthermore, anisotropic expansion induces a linear reflection between outward and inward modes, governed by the matrix $\mathbb M = \text{diag}\left (\frac{3}{4} \gamma_w - \frac{1}{2}\gamma_a, - \frac{1}{4}\gamma_w + \frac{1}{2} \gamma_a,  - \frac{1}{4}\gamma_w + \frac{1}{2} \gamma_a,\right )$. 
In the non-expanding, non-accelerating case ($a = 1, w = 1$), the system recovers the standard non-expanding compressible MHD equations, as described by \citet{Marsch1987}. Furthermore, for studies focused strictly on Alfvénic turbulence, the equations can be evaluated in the incompressible limit ($\nabla \cdot \mathbf{v} = 0, \nabla \rho = 0$), where the system is described through Equation \eqref{elsasser2} as
\begin{equation}
    \begin{aligned}[b]
         \frac{D^\mp z^\pm}{D t} = \pm \frac{1}{4} (\mathbf z^+ - \mathbf z^-) \text{Tr}(\mathbb H)\\-  \frac{1}{8}  \nabla (\mathbf z^+ - \mathbf z^-)^2 - \frac{1}{4} \text{Tr}(\mathbb H) \mathbf z^\pm - \mathbb M \cdot \mathbf z^\mp.\label{incompressible}
    \end{aligned}
\end{equation}
It is relevant to note that the symmetry in this generalized Els\"asser formulation relies entirely on the covariant correction of the ideal, expanding MHD system, as additional expansion matrices are simplified in our modified Faraday´s Law. Furthermore, this framework is the Lagrangian counterpart to stationary Eulerian transport models (Equation (32) in~\citet{Cranmer_2005} and Equation (1) in~\citet{Verdini_2007}), extending this formalism to include compressible effects in Equation \eqref{elsasser2}. In our framework, the trace of the expansion rate matrix, $\text{Tr} (\mathbb H)$, naturally accounts for the WKB amplitude attenuation of waves due to plasma expansion, while $\mathbb M$ acts as a non-WKB linear reflection source. 

In standard non-WKB theory, this linear reflection is driven by gradients in the solar wind average fields, especially at low frequencies~\citep{Heinemann1980, Zhou_1989}.  Previous Accelerating Expanding Box (AEB) models \citep{Tenerani2013, Nariyuki2015, Tenerani2017} rely on classical coordinate transformations where velocity and density gradients are retained as external source terms to drive wave reflection. In contrast, our covariant deduction absorbs these gradients into the box geometry, showing that in this formalism, wave reflection emerges as a consequence of plasma expansion and acceleration itself, isolated in the equations through the $\mathbb M$ matrix term. This formulation provides a cleaner theoretical description and a significant computational advantage by simplifying the numerical integration of the system. 

\section{Wave Analysis and Numerical Implementation}
\label{sec5}

\subsection{Dispersion Relation Analysis}

To validate our theoretical framework and show the utility of our set of equations to capture non-WKB effects, this section presents a numerical integration of Equations \eqref{elsasser2} and \eqref{incompressible} to study wave propagation and reflection in a linear regime. These equations can be solved to obtain the propagation and dispersion of Alfvén and sound modes in a linear, small-amplitude limit. 
Following the standard procedure for deriving linear plasma wave equations~\citep{Stix1992}, we consider small-amplitude perturbations to the $\delta z^\pm$ fields and plasma quantities.

We linearize the system such that $\mathbf z^\pm = \mathbf z_0^\pm + \delta \mathbf z^\pm,$, where $\mathbf u = \delta \mathbf u$ and $\mathbf b = \mathbf b_0 + \delta \mathbf b$. Thus, $\mathbf z_0 = \pm \mathbf b_0$, and $\mathbf z^\pm = \pm \mathbf b_0 + \delta \mathbf z^\pm$. On the other hand, $\rho = \rho_0 + \delta \rho$ and $P_T = P_{T0} + \delta P_T$. Here, the background quantities are spatially homogeneous, such that $\nabla \cdot \mathbf b_0 = 0$ and $\nabla P_{T0} = 0$, and perturbations are sufficiently small. We consider $\phi_{0} \gg \delta \phi$ for each quantity $\phi$. With this, Equation \eqref{elsasser2} can be written as the equation for the MHD modes within the EBM model as

\begin{equation}
\begin{aligned}[b]
     \frac{\partial \delta \mathbf z^\pm}{\partial t} \mp (\mathbf b_0 \cdot \nabla) \delta \mathbf z^\pm = \pm \frac{1}{2} \mathbf b_0 \left (\frac{\partial}{\partial t} \pm (\mathbf b_0 \cdot \nabla) \right ) \frac{\delta \rho}{\rho_0} \\- \nabla (\mathbf b_0 \cdot \delta \mathbf b) -  \left ( \tilde v_S^2 + \frac{1}{2} \mathbf b_0^2\right )\nabla \frac{\delta \rho}{\rho_0} \\-  \frac{1}{4} \text{Tr}(\mathbb H) \delta \mathbf z^\pm - \mathbb M \cdot \delta \mathbf z^\mp. \label{waveeq}
\end{aligned}
\end{equation}

Equation \eqref{waveeq} describes the coupled evolution of magneto-acoustic and Alfvénic fluctuations in the expanding solar wind. The last two terms on the right-hand side introduce the effects of plasma expansion as non-inertial forces, representing geometric damping and expansion-induced mode-coupling. To present clear dispersion relations of MHD normal modes, we decouple the general equation. By imposing the incompressible ($\delta \rho = 0$) and transverse magnetic field fluctuations ($\mathbf b_0 \cdot \delta \mathbf b = 0$) limits in this configuration, we isolate the dynamics of pure Alfvén waves. On the other hand, imposing $\delta \mathbf b = 0$ removes the magnetic tension terms, isolating the longitudinal acoustic modes.

\begin{figure}[h]
    \centering
    \includegraphics[width=0.8\linewidth]{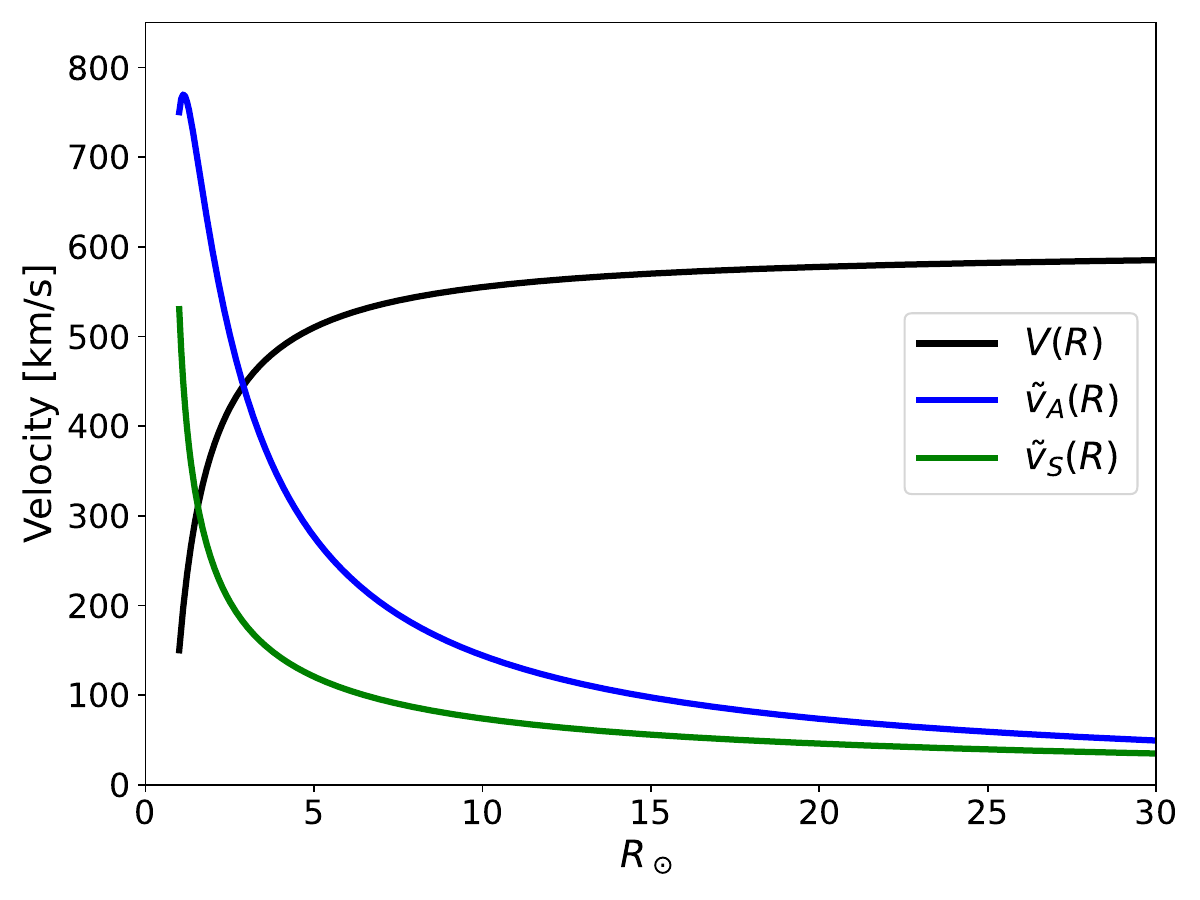}
    \caption{Solar wind velocities in our EBM framework as a function of heliocentric distance in solar radii $R_\odot$: Solar wind speed $V(R)$, Alfvén speed $\tilde v_A(R)$, and sound speed $\tilde v_s(R)$ as given by the EBM formalism.}
    \label{fig:vel_profiles}
\end{figure}

\begin{figure*}[ht!]
    \centering
    \includegraphics[width=0.7\linewidth]{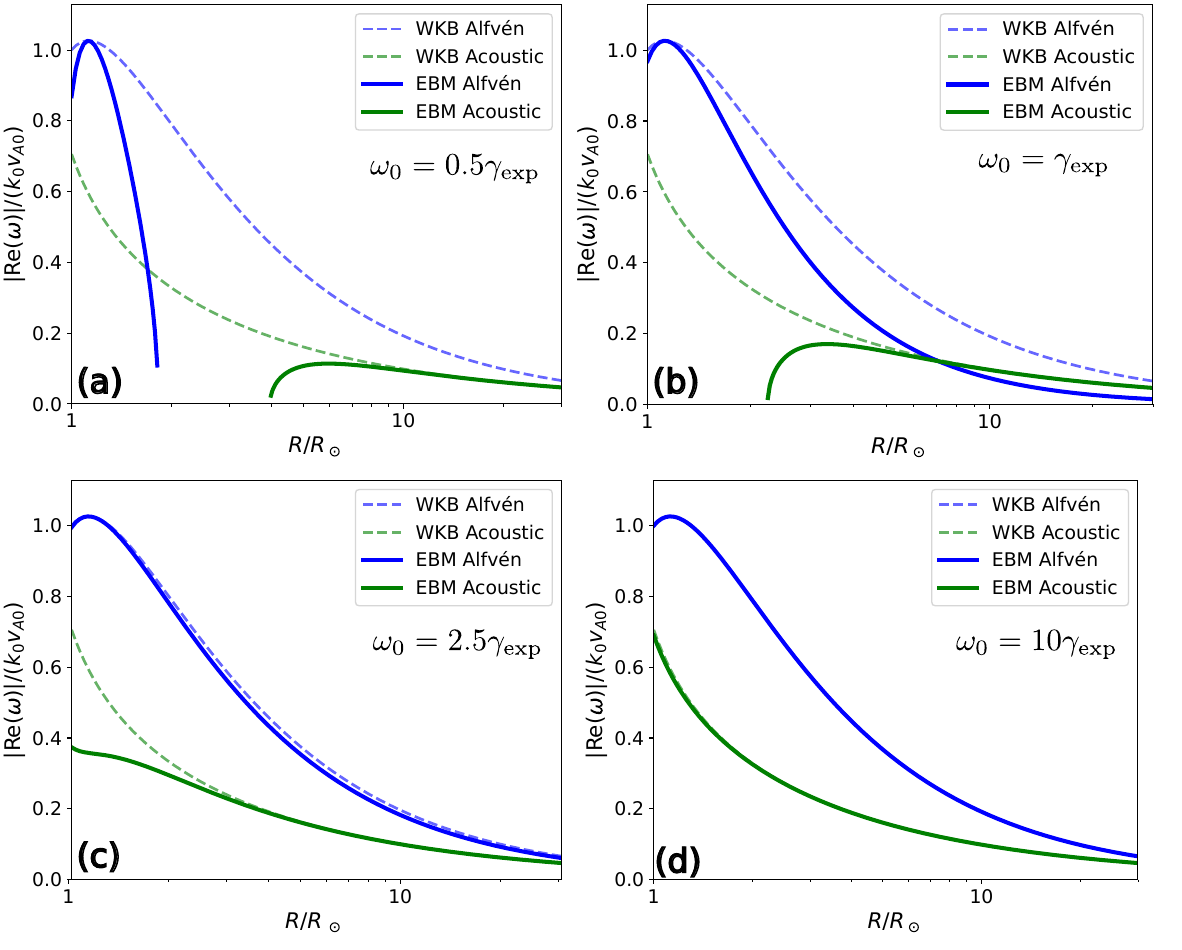}
    \caption{Radial evolution of the real part of the complex dispersion relations, normalized by the initial Alfvén frequency $\omega_0 = k_0 v_{A0}$,  obtained from Equations \eqref{alfvenw}-\eqref{soundw}. The fixed initial wavenumber $k_0$ is expressed in terms of initial frequencies, which are multiples of the initial characteristic expansion rate $\gamma_{\text{exp}} = V_0/R_0$. Panels (a)--(d) display the real frequencies $|\text{Re}(\omega)|$ as a function of heliocentric distance $R/R_\odot$ on a semi-logarithmic scale for: (a)-(b) a low-frequency case ($\omega_0 = 0.5 \gamma_{\text{exp}}$, $1.0 \gamma_{\text{exp}}$ ) showing the strong cutoff in the non-WKB limit, (c)
    a transition regime ($\omega_0 = 2.5 \gamma_{\text{exp}}$) showing a convergence to the WKB limit further from the Sun, and (d) a high-frequency case ($\omega_0 = 10 \gamma_{\text{exp}}$) representing the WKB limit.}
    \label{fig:dispersion}
\end{figure*}

We follow standard linear theory~\citep{Stix1992} to obtain the wave dispersion relations. We assume a radial background field $\mathbf{B}_0 = B_0 \hat{x}$ and restrict our analysis to parallel-propagating modes ($\mathbf{k} = \tilde k \hat{x}$), and thus apply a local Fourier ansatz proportional to $\exp[i(kx - \omega t)]$ to the isolated modes of Equation \eqref{waveeq}, where $\omega$ and $k$ are the wave frequency and wavenumber, respectively. In a uniform plasma, this transformation yields an algebraic dispersion relation. However, in the EBM framework, the expansion rates, contained in $\text{Tr}(\mathbb H)$ and $\mathbb M$, introduce explicit time dependencies. Traditionally, this prevents a direct Fourier analysis, forcing previous analytical works to rely on a strict WKB approximation that neglects the effects of expansion~\citep{Saldivia2025}. To overcome this limitation, we algebraically absorb the non-inertial terms into a set of generalized complex eigenfrequencies ($\Omega_A = \omega_A + i \text{Tr}(\mathbb H)/4jkj$ for the Alfvén mode and $\Omega_S = \omega_S + i \frac{\dot w}{2w}$ for the acoustic mode). By maintaining the non-inertial forces in the imaginary plane, solar wind expansion acts as a geometric damping rate. This approximation implies assuming that the temporal derivatives of the expanding terms are smooth with respect to the wave timescale, instead of neglecting these expansion terms. For the Alfvén and acoustic modes, respectively, this generalization yields the dispersion relations $\omega_A = \pm \sqrt{k^2 \tilde v_A^2 - \mathbb M_\perp^2} -   i\frac{1}{4} \text{Tr} (\mathbb H)$  for the Alfvén mode, and $\omega_S = \pm \sqrt{k^2 \tilde v_S^2 - \frac{\gamma^2_w}{4}} - i \frac{\gamma_w}{2}$ for the sound mode. We normalize these expressions as
\begin{equation}
    \frac{\omega_A}{k_0 v_{A0}}  = \pm \sqrt{\frac{1}{a^2} - \frac{\mathbb M_\perp^2}{k_0^2 v_{A0}^2}} - i \frac{1}{4} \frac{(2 \gamma_a + \gamma_w )   }{k_0 v_{A0}},\label{alfvenw}  
\end{equation}
\begin{equation}
\frac{\omega_S}{k_0 v_{A0}} = \pm \sqrt{ \frac{\beta(R)}{a^2}  - \frac{\gamma_w^2}{4 k_0^2 v_{A0}^2} } - i \frac{\gamma_w}{2 k_0 v_{A0}},\label{soundw}
\end{equation}
where $\mathbb M_\perp = -\frac{1}{4} \gamma_w + \frac{1}{2} \gamma_a$ is the perpendicular component of the reflection matrix and $\beta(R) = \beta_0 \, w^{-\Gamma} a^{4 - 2\Gamma}$, where $\beta_0 = v_{S0}^2/v_{A0}^2$ is the initial plasma beta. Here, $\tilde v_A \propto w^{1/2}  a^{-1}$ is the time-dependent Alfvén speed, as from Equation \eqref{eq_cont_s} at background order, the radial profile of background density is $\rho_0 \propto (w a^2)^{-1}$, while the radial magnetic field scales as $B_0(a) \propto a^{-2}$. The physical wavenumber is $\tilde {k}  = k_0/w$, with $k_0$ the initial wavenumber due to the longitudinal stretching of the plasma parcel. A detailed derivation of these radial profiles for the non-accelerating case can be found in \citet{Saldivia2025}. This confirms that our covariant deduction is consistent with standard local MHD theory~\citep{Grappin1993, Saldivia2025}, where Alfvén waves remain an exact solution of the EBM-MHD system. The appearance of the $w^{1/2}$ factor is a direct consequence of the plasma expansion. While transverse expansion represented by $a$ affects both the mass density and the magnetic field, the longitudinal acceleration in $w$ only affects the mass density ($\rho_0 \propto a^{-2}w^{-1}$, $B_0 \propto a^{-2}$). As a consequence, in this framework, longitudinal stretching can increase the local Alfvén speed.

To numerically model this dispersion relation, we need to define the evolution of the expansion rates $a$ and $w$. In order to accurately capture near-Sun solar wind conditions, we adopt an empirical asymptotic velocity profile. Observations and fluid models show that the solar wind accelerates rapidly in the lower corona before reaching a nearly constant terminal velocity at around $\sim 10 R_\odot$~\citep{Parker1958, Sheeley1997, Verscharen2019}. We model the velocity profile parameterized in terms of the asymptotic terminal velocity $V_\infty$, assuming a steepness exponent $\alpha = 1$ as~\citep{Knigge1995,Hagino2015}
\begin{equation}
    V(R) = V_0 + (V_\infty - V_0)\left (1 - \frac{R_0}{R} \right ).\label{velprofile}
\end{equation}
This model translates conveniently into the EBM framework. Since $a = R/R_0$, the longitudinal acceleration factor $w = V(R)/V_0$ can be written in terms of $a$ as $w(a) = 1 + ( {V_\infty}/{V_0} - 1) \left(1 - 1/a\right)$. Figure \ref{fig:vel_profiles} shows the radial profiles employed in this work for the solar wind speed $V(R)$, and plasma Alfvén $\tilde v_A(R)$ and sound $\tilde v_S(R) $ speeds. For our analysis, we consider typical fast solar wind conditions at the coronal base, setting an initial Alfvén speed $v_{A0} = 750$ km s$^{-1}$, initial flow speed $V_0 = 150$ km s$^{-1}$, terminal speed  $V_\infty = 600$km/s, an initial plasma beta $\beta_0 =  0.5$, and a base radius $R_0 = 1.0 R_\odot$. Under the assumption of a stationary background flow, the expansion rates $\gamma_w$ and $\gamma_a$ become explicit functions of the normalized distance $R/R_0$. Thus, for our numerical integration, we parameterize all time dependencies in our equations as functions of the expansion parameter $a = R/R_0$ instead of time $t$. This spatial parameterization is analytically equivalent to the temporal integration, as shown in Appendix \ref{appendixC}.  We also consider $R_0 = 1.0 R_\odot$.

\begin{figure*}[ht!]
    \centering
    \includegraphics[width=0.75\linewidth]{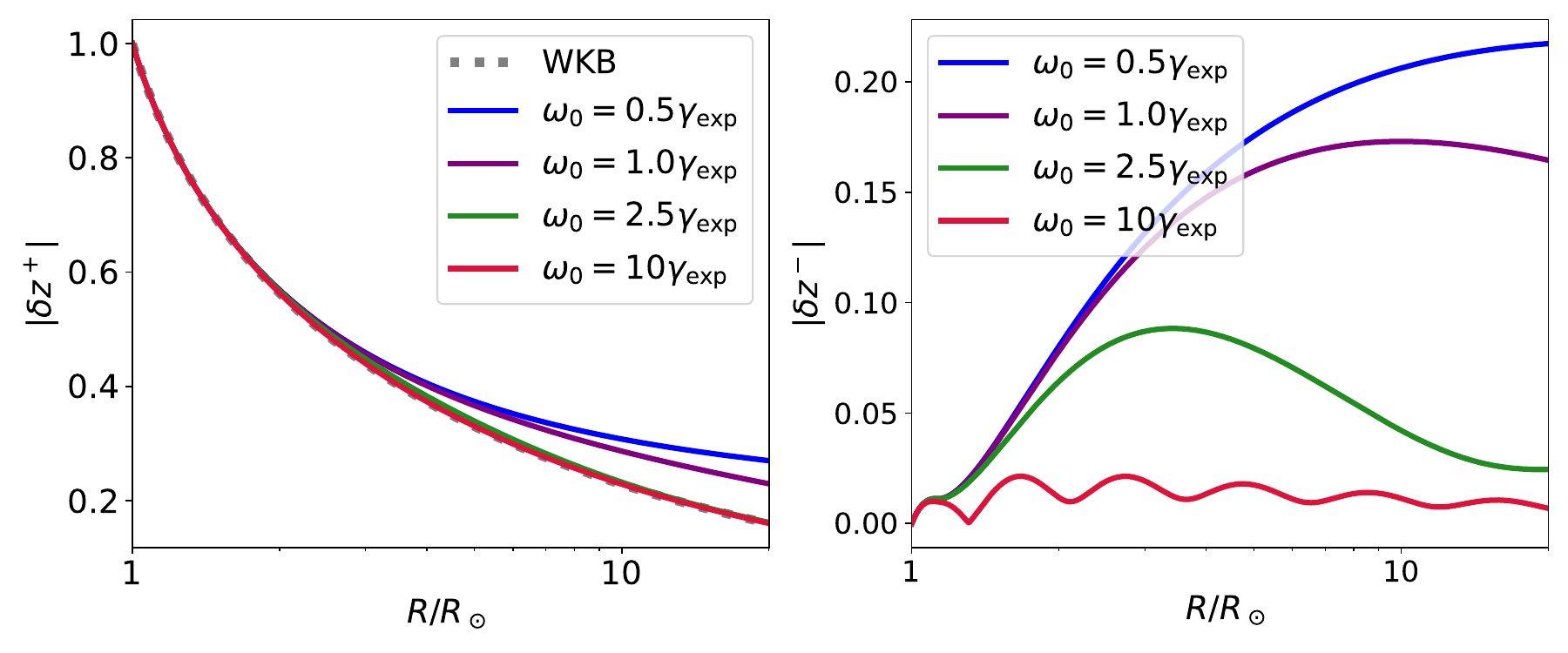}
    \caption{Radial evolution of the Els\"asser amplitudes ($|\delta z^+|, |\delta z^-|$) for an initially purely outward-propagating wave. Left panel: Amplitude evolution of the primary forward-propagating mode $|\delta z^+|$ for four different initial frequencies ($\omega_0 = 0.5 \gamma_{\text{exp}}$, $\omega_0 = 1.0 \gamma_{\text{exp}}$, $\omega_0 = 2.5 \gamma_{\text{exp}}$, $\omega_0 = 10 \gamma_{\text{exp}}$), as a function of heliocentric distance $R/R_\odot$. The black dotted line represents the theoretical WKB  envelope. Right panel: Amplitude evolution of the reflected mode $|\delta z^-|$ for the same set of initial frequencies as a function of heliocentric distance $R/R_\odot$.}
    \label{fig:amplitudes}
\end{figure*}

Figure \ref{fig:dispersion} shows the radial evolution of the normalized complex dispersion relations, derived from Equations \eqref{alfvenw} and \eqref{soundw}, evaluated for several fixed initial wavenumbers, written in terms of the initial frequency $\omega_0 = k_0 v_{A0}$. This initial frequency is expressed as multiples of the characteristic expansion frequency at the moment the wave is generated. As such, we define $\gamma_a(R_0) = \gamma_\text{exp} = V_0/R_0$. Here, we present the radial evolution of the real part of the frequencies $|\text{Re}(\omega)|$, which illustrates the oscillatory nature of the modes, with solid lines representing the EBM solutions and dashed lines indicating the classical WKB limits.  

For the lowest frequency mode $(\omega_0 = 0.5 \gamma_{\text{exp}},$ panel a), the real frequency collapses to zero across significant spatial regions, suggesting that solar wind expansion and acceleration impose a frequency threshold. For this mode frequency, the wave is totally reflected and ceases to propagate in that spatial range. This implies that the wave energy cannot be emitted as traveling waves in the interplanetary medium, suggesting that while high-frequency waves escape to heat the extended heliosphere, non-propagating low-frequency modes are subject to significant wave reflection. Panel (b) shows a low-frequency mode comparable to the expansion rate $(\omega_0 = \gamma_{\text{exp}})$. This mode shows significant wave reflection on the acoustic mode near the Sun. On the other hand, the Alfvén mode is able to propagate, although showing significant deviation from the WKB limit in the mode frequency. 

The transition regime ($\omega_0 = 2.5 \gamma_{\text{exp}}$, panel c) reveals a noticeable departure from the WKB limit near the coronal base for the acoustic mode. Both modes overcome the cutoff barrier. At larger heliocentric distances, the acoustic mode frequency converges to the WKB limit. For the high-frequency limit ($\omega_0 = 10 \gamma_{\text{exp}}$, panel d), the EBM frequencies perfectly overlap with the classical WKB approximation. This indicates that the inertial expansion effects are negligible compared to wave propagation for this mode frequency ($k^2 \tilde v_{A, S}^2 \gg \mathbb M_\perp^2, \gamma_w^2/4$), implying that plasma expansion plays a negligible role in the evolution of high-frequency modes. It is worth noting that, under the parameters defined in our velocity profile, the initial frequencies correspond to $\omega_0 \sim 10^{-4}$ rad/s, $\sim 2 \times 10^{-4}$ rad/s, $\sim 5 \times 10^{-4}$ rad/s, and $ \sim 2 \times  10^{-3}$ rad/s, respectively.

Furthermore, a relevant departure from classical theory is revealed in Equations \eqref{alfvenw}-\eqref{soundw}. While the standard WKB approximation relies on strictly real frequencies, our EBM formulation captures the expansion-driven attenuation as imaginary frequency components $|\text{Im}(\omega)|$. These geometric damping rates, dictated by the non-inertial terms $\gamma_{\text{exp}}$ and $\gamma_w$, are independent of the initial frequency $\omega_0$. This wave attenuation reaches its maximum near the Sun, where gradients are steepest~\citep{Heinemann1980, Zhou_1989}, and vanishes at larger heliocentric distances.

It is relevant to note that relying solely on a local dispersion relation assumes a timescale separation approximation, requiring the temporal variation of the expanding rates to be slower than the wave variation timescales. However, for frequencies low compared to the expansion frequency $(\omega = 0.5 \gamma_\text{exp}, 1.0 \gamma_\text{exp})$, this approximation is challenged due to the steep velocity profile. Precisely because of this local WKB breakdown, simple analytical approximations are insufficient, necessitating numerical integration of the full differential system.

\subsection{Wave Amplitude and Reflection Analysis}

To further validate our analytical results and explore the breakdown of the WKB limit, we numerically integrate the linear time-dependent evolution equation \eqref{waveeq} for the isolated Alfvénic modes in the incompressible limit ($\delta \rho = 0, \delta u_x = 0$). The system is integrated in the co-moving frame using a fourth-order Runge-Kutta (RK4) method to track the evolution of the Els\"asser variables ($|\delta z^+|, |\delta z^-|$) and capture the expansion-induced wave reflection and mode coupling. 

Figure \ref{fig:amplitudes} shows the radial evolution of these mode amplitudes. The left panel shows the amplitude evolution for the primary forward-propagating mode ($|\delta z^+|$) for different $\omega_0$ values. This allows us to compare the numerical decay directly against the analytical envelopes dictated by the WKB limit. The outward amplitude ($|\delta z^+|$) follows the geometric decay of classical WKB waves ($\propto (a^2 w)^{-1/4}$) almost entirely at higher frequencies, while this decrease is slighter than in the WKB limit for the low-frequency waves, suggesting that plasma expansion introduces significant non-WKB effects at low frequencies compared to the expansion rate, with these effects being negligible at higher frequencies. 

\begin{figure*}[ht!]
    \centering
    \includegraphics[width = 0.8\linewidth]{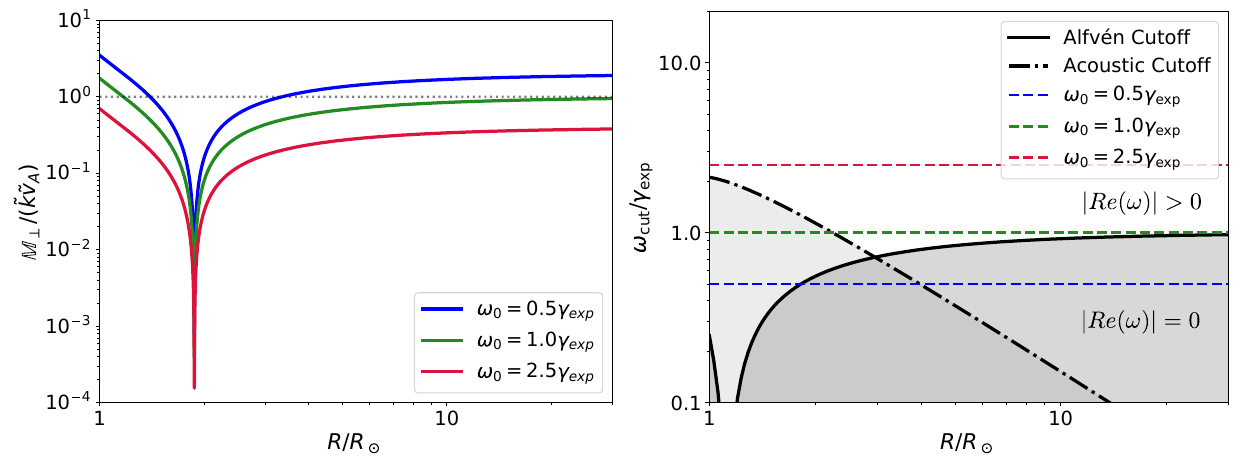}
    \caption{Breakdown of the WKB approximation and expansion-driven Alfvén wave reflection in the expanding solar wind. Left panel: Radial evolution of the local reflection parameter $\mathbb{M}_\perp / (\tilde k \tilde{v}_A)$ for three initial frequencies. The black dotted line at unity represents the WKB limit. Frequencies above this limit enter the regime where the reflection matrix dominates over wave propagation. Right panel: Radial profile of the critical cutoff frequency $\omega_{\text{cut}}$, where $\mathbb M_\perp = \tilde k \tilde v_A$, normalized by the expansion frequency $\gamma_{\text{exp}}$. The solid black curve shows the boundary between the wave propagation regime ($|\text{Re}(\omega)| > 0$) and the non-propagating zone ($|\text{Re}(\omega)| = 0$, shaded gray). Horizontal dashed lines denote the initial frequencies evaluated in the left panel.}
    \label{fig:reflection_coeff}
\end{figure*}

On the other hand, the right panel of Figure \ref{fig:amplitudes} shows the dependence of the reflected mode amplitude $|\delta z^-|$ on the initial frequency $\omega_0$. In the high-frequency limit ($\omega_0 = 10 \gamma_{\text{exp}}$), the wave-propagation term suppresses the background expansion gradients ($\tilde {k} \tilde {v} _ A \gg \mathbb {M} _ \perp$), avoiding significant reflection and recovering the classical WKB approximation. As the frequency decreases to the transition regime ($\omega_0 = 2.5 \gamma_{\text{exp}}$), plasma expansion drives the aforementioned localized reflection peak. For the lower frequency modes ($\omega_0 = 0.5 \gamma_{\text{exp}}$, $1.0 \gamma_{\text{exp}}$), the reflected mode exhibits a considerable growth across the spatial domain. In this low-frequency regime, plasma expansion drives a significant reflection of modes further from the Sun. As the plasma expansion frequency exceeds the wave frequency, wave propagation is suppressed, thereby forcing continuous energy transfer from the outward mode to the reflected mode across the inner heliosphere.

This non-WKB reflection, driven by the transverse coupling matrix $\mathbb M_\perp$, grows to a peak in each case before decaying asymptotically, showing the mode reflection induced by the solar wind expansion. Although this reflected mode arises in a linear model, the presence of these counter-propagating waves is an essential prerequisite for the origin of non-linear turbulent cascades in the solar wind~\citep{Zhou_1989, Matthaeus_1999}, suggesting that solar wind expansion and acceleration may play a role in the origin of the turbulent cascade, primarily for lower frequencies compared to the expansion rate.

Furthermore, to quantify the spatial range over which this wave reflection occurs, Figure \ref{fig:reflection_coeff} presents the breakdown of the WKB approximation in the expanding solar wind. The left panel shows the radial evolution of the local Alfvénic reflection parameter $\mathbb M_\perp = -\frac{1}{4} \gamma_w + \frac{1}{2}\gamma_a$ normalized to the local, co-moving WKB Alfvén frequency $\tilde k \tilde v_A(R)$. The threshold for WKB validity is marked by the black dotted line at unity. When a wave crosses this threshold, the plasma expansion overcomes the wave propagation, causing reflection, as is the case for the only the lowest frequency mode ($\omega_0 = 0.5 \gamma_{\text{exp}}$). For $\omega_0 = 1.0 \gamma_{\text{exp}}$, the curve converges to unity at larger heliocentric distances, showing that this frequency is the upper limit over which significant reflection occurs for the Alfvénic mode. Higher-frequency modes ($\omega_0 = 2.5 \gamma_{\text{exp}}$ ) remain below this limit throughout the domain, where no significant wave reflection occurs. It is worth noting that, for all frequencies, the reflection coefficient diverges to 0 in a particular spatial range where $\frac{1}{4} \gamma_w = \frac{1}{2} \gamma_a$, where WKB effects are recovered. This critical heliocentric distance can be analytically quantified in our velocity profile as
\begin{equation}
    R_c = \frac{3}{2}R_0 \left ( 1 - \frac{V_0}{V_\infty}\right ). 
\end{equation}
This critical radius $R_c$ represents a localized spot in the inner heliosphere where the transverse expansion rate of the solar wind balances out the acceleration rate. At this location, the wave experiences no stretching, suppressing mode reflection induced by plasma expansion.

On the other hand, the right panel of Figure \ref{fig:reflection_coeff} displays the radial profile of the critical cutoff frequency $\omega_{\text{cut}}$, where the reflection parameter equals unity ($\mathbb M_\perp = \tilde k \tilde v_A$). The solid black curve separates the wave propagation regime ($|\text{Re}(\omega)| > 0$) from the non-propagating, reflection-dominated zone ($|\text{Re}(\omega)| = 0$, shaded gray), for Alfvén waves. The dashed-dotted line separates the regimes for acoustic modes. By superimposing the initial frequencies as horizontal dashed lines, we show that the lowest-frequency mode is in the evanescent zone near the solar corona, showing the collapse of its real frequency and the generation of the backward-propagating mode seen in our numerical integration. For $\omega_0 = 1.0 \gamma_{\text{exp}}$, the wave frequency converges to the cutoff limit at larger heliocentric distances for the Alfvénic mode, which propagates freely, but the acoustic mode is subject to significant reflection near the Sun. Finally, the transition frequency $\omega_0 = 2.5 \gamma_{\text{exp}}$ and higher remain within the propagation regime across all heliocentric distances. Note that this cutoff frequency $\omega_{\text{cut}}$ defines a threshold for wave propagation in the lower corona.

\section{Discussion and Conclusions}
\label{sec6}

In this work, we revisited the Expanding Box Model (EBM) equations from first principles to address a long-standing structural limitation in the magnetohydrodynamic modeling of the solar wind. Historically, the EBM has relied on coordinate transformations that transform the MHD equations to a co-moving grid, while implicitly treating the magnetic field as a Galilean invariant across reference frames $(\mathbf B = \mathbf B')$~\citep{Grappin1996,Echeverria-Veas2023}. To overcome this approximation, we modeled the solar wind expansion as a geometric property of spacetime itself, utilizing an anisotropic Bianchi Type I spacetime metric. By deducing the relativistic and non-relativistic ideal MHD equations directly from covariant conservation laws, we coupled the plasma and the electromagnetic fields to the expanding macroscopic background, in contrast to previous formulations. 

Recent deductions of the EBM equations from first principles have highlighted a theoretical asymmetry that originates in the earliest formulations. While historical works correctly recognized that vector fields such as the plasma velocity must be geometrically scaled to account for macroscopic stretching~\citep{Liewer2001, Innocenti_2019}, they treated the magnetic field as an exception to this coupling between vectors and expanding space. Thus, they generated an asymmetric vector transformation. This is the direct mathematical cause of the expansion matrices trapped within the magnetic operators, such as the co-moving magnetic tension written as $[\mathbf B \cdot (\mathbb A^{-1} \cdot \nabla)](\mathbb A \cdot \mathbf v)$ in \cite{Echeverria-Veas2023}. Our covariant framework resolves this asymmetry by treating all vector fields as proper geometric tensors. Thus, the vector scaling matrices cancel out the inverse gradient terms.  This exact algebraic cancellation restores the symmetry of the ideal MHD equations within the co-moving domain.

Furthermore, we also demonstrated that eliminating these algebraic artifacts from the co-moving frame $S'$ does not alter the physical observables in the inertial frame $S$. By transforming our co-moving equations into the physical, heliospheric inertial frame, we recovered the exact MHD equations established in the literature.
Our geometric derivation yields the correct non-inertial source terms required for the Accelerating Expanding Box (AEB) regime, coinciding with the vector formulations of \citet{Tenerani2013} and \citet{Nariyuki2015} in the limit where longitudinal acceleration is present ($\dot w(t') \neq 0$). Furthermore, we demonstrated that the macroscopic Parker spiral structure of the interplanetary magnetic field, which arises within the EBM framework, is a purely geometric projection of expanding spacetime, rather than a local property of the plasma.  Within the fully covariant co-moving frame $S'$, all components of the magnetic field naturally decay isotropically $(\mathbf B' \propto 1/a^2)$. 

Because this covariant deduction does not rely on \textit{a priori} coordinate transformations, the macroscopic scale factors $a$ and $w$ remain completely general. Consequently, the resulting equations allow for diverse regimes of expansion. Complex physical profiles, such as non-uniform radial acceleration, asymptotic velocity profiles, or acceleration in the transverse directions, can be incorporated without altering the mathematical structure. This geometric framework not only refines the modeling of the solar wind but also provides a versatile foundation for exploring broader astrophysical phenomena.

Ultimately, this covariant formulation provides a mathematically consistent framework in which the rigorous transformation of the magnetic field yields several theoretical and observational implications for solar wind modeling. While compatible with the Galilean approximation, this approach incorporates the correct coordinate transformations into the model, thus eliminating the geometric residues and time-dependent expansion factors from the spatial differential operators in previous formulations~\citep{Echeverria-Veas2023}. This formulation clarifies the physical distinction between the co-moving and inertial frames, offering a new tool for observational studies by allowing the study of plasma dynamics as a function of measurement location. This has clear observational implications, allowing for the direct application of theoretical expanding models to \textit{in situ} heliospheric observations for a spacecraft co-moving with the solar wind.


As a step toward connecting this theoretical framework with potential computational applications on MHD turbulence in the expanding solar wind, we presented the EBM-MHD system in terms of Els\"asser variables. This generalized system provides a rigorous framework for studying compressible turbulence under the influence of plasma acceleration and expansion. Notably, for numerical implementations, directly evolving $\mathbf{z}^\pm$ allows simulation codes to accurately track the turbulent cascade while incorporating radial plasma expansion and acceleration profiles. The explicit appearance of the linear mode-coupling term $-\mathbb{M} \cdot \mathbf{z}^\mp$ introduces a macroscopic interaction between inward- and outward-propagating fluctuations. This mathematical structure opens the door to studying whether plasma expansion acts as a continuous source for induced wave reflection. Exploring how this geometric coupling might affect the evolution of cross-helicity in the inner heliosphere remains an open problem for future MHD turbulence models.

The dispersion-relation analysis of the linearized EBM-MHD system reveals that the spatial gradients of the expanding solar wind, in our formalism, manifest as damping rates ($|\text{Im}(\omega)|$) and impose a low-frequency cutoff. This was corroborated through RK4 amplitude integration, where we showed that the reflection matrix $\mathbb{M}_\perp$ drives a localized, non-WKB reflection of Alfvén waves. Furthermore, we identified the existence of a critical heliocentric distance $R_c$, where transverse expansion balances longitudinal acceleration to momentarily restore WKB propagation. We showed that these non-WKB effects driven by expansion introduce significant corrections to the wave frequencies for fluctuations with a frequency similar in magnitude to the expansion frequency ($\omega_0 = 0.5 \gamma_{\text{exp}}$). For the velocity profile we considered, this corresponds to fluctuations near the energy injection range ($\sim 10^{-4}$ rad/s), while frequencies in the inertial range ($\omega_0 \sim 10^{-3}$ rad/s) propagated in the WKB limit. It is well established that steep gradients in the solar wind are relevant in inducing mode reflection, especially for frequencies low compared to the expansion frequency \citep{Heinemann1980, Zhou_1989, Matthaeus_1999}. Consequently, this work shows that our covariant EBM is able to capture these non-WKB effects on MHD wave propagation, confirming that solar wind expansion and acceleration may play a major role in this process. It is relevant to note that these analytical results rely on a local time-scale separation approximation, with a validity that depends on the chosen velocity profile. Because the acceleration near the Sun challenges this local approximation for low-frequency modes, future work will focus on relaxing this constraint, either by implementing more self-consistent velocity models or by relaxing the time-scale separation in the EBM equations.

Our results rely on a linear approximation, as in the considered inner-heliospheric domain (up to $\sim 15 R_\odot$), the transit time of the outward-propagating Alfvén waves is considerably shorter than the wave period for the lowest frequencies. As a consequence, the waves traverse the spatial range before nonlinear effects are significant in the wave dynamics. However, to accurately describe the wave evolution at larger heliospheric distances, it becomes necessary to incorporate nonlinear interactions. A non-linear approach could modify the predicted population of $z^-$ modes generated and sustained in the interplanetary medium. Incorporating such dynamics through approaches such as truncated models \citep{Dmitruk_2001} is left for future work. Nevertheless, our formulation already allows us to compare our linear predictions of the $z^-$ population with in-situ observations from missions such as Parker Solar Probe, allowing to quantify the competition between reflection induced by solar wind expansion and non-linear interactions.

On the other hand, the frequency dependence of this expansion-driven mode reflection can contribute to the larger problem of the origin of solar wind turbulence. Our results show that solar wind expansion acts as an efficient generator of counter-propagating Alfvén waves ($|\delta z^-|$), primarily for low-frequency modes ($\omega_0 \lesssim \gamma_{\text{exp}}$). Conversely, higher frequencies become increasingly decoupled from solar wind expansion, suppressing linear reflection and converging toward the classical WKB limit. This low-frequency reflection mechanism complements the non-linear dynamics extensively described in the literature. \citet{Shoda_2018} demonstrated that at frequencies $\omega_0 \gtrsim 10^{-3}$ Hz, Alfvénic fluctuations are subject to the Parametric Decay Instability (PDI), a wave-wave interaction widely recognized as a primary mechanism for the generation of backward-propagating modes at smaller scales \citep{Tenerani2013, Del_Zanna2015,Saguchi_2026}. Remarkably, this frequency range coincides with the threshold where our expansion-induced linear reflection becomes less relevant in Alfvénic fluctuations. While further studies are required to couple these spatial and temporal scales, this overlap suggests that while solar wind expansion might act as the primary linear source of counter-propagating waves at large scales, non-linear processes such as PDI take over to sustain the turbulent cascade across the higher-frequency spectrum.

\begin{acknowledgments}
    We are grateful for the support of ANID Chile through the National Doctoral Scholarships No. 21250323 (SS), and FONDECYT grants No. 1240281 (PSM) and 1230094 (FAA). We also acknowledge DICYT Project 042531SSSA\_Postdoc (SEV).
\end{acknowledgments}

\appendix

\section{EBM Metric}
\label{appendixA}

\subsection{Expanding Geometry and Frame Transformations}
\label{appendixa1}
The invariant physical distance measured from within the co-moving frame $S'$ is described by a Bianchi Type I metric, used to model anisotropic space expansion~\citep{Taub1951, MTW}. The line element is given by
\begin{equation}
    ds^2 = - dt'^2 + w^2(t') dx'^2 + a^2(t') dy'^2 + a^2(t') dz'^2.\label{metric}
\end{equation}
From this line element, the corresponding metric tensor $g_{\mu'\nu'}$, which represents how distances and dot products are evaluated within this stretching coordinate system, is
\begin{align}
    g_{\mu' \nu'} &= \text{diag}(-1,w^2,a^2,a^2), \\ g^{\mu' \nu'} &= \text{diag}(-1,w^{-2},a^{-2},a^{-2}),
\end{align}
where the metric determinant is $g = -w^2 a^4$. The non-zero Christoffel symbols of the second kind $\Gamma_{\alpha\beta}^\mu$ are symmetric in this system, and given by
\begin{align}
\Gamma_{t'x'}^{x'} &= \frac{\dot w}{w}, & \Gamma_{t'y'}^{y'} &= \frac{\dot a}{a}, & \Gamma_{t'z'}^{z'} &= \frac{\dot a}{a}, \nonumber \\
    \Gamma_{x'x'}^{t'} &= w \dot w, & \Gamma_{y'y'}^{t'} &= a \dot a, & \Gamma_{z'z'}^{t'} &= a \dot a. \label{christoffel}
\end{align}
Note that the $\Gamma_{t' ,i'}^{i'}$ factors are inversely proportional to the characteristic expansion time scales of the solar wind, and correspond to the fictitious inertial forces arising from the expansion.

To relate the vector fields in the co-moving frame $S'$ to the measurable physical quantities in a locally flat, inertial laboratory frame $S$, we utilize an orthonormal tetrad field. The tetrad basis vectors $\mathbf{e}_{\mu'}$ are the coordinate basis vectors inside the expanding box. As these are not unitary, their lengths are directly determined by the metric tensor ($\|\mathbf{e}_{i'}\| = \sqrt{g_{i'i'}}$), associating the expanding frame with the inertial frame as
\begin{align}
    \mathbf e_{t'} = \mathbf e_{t}, && \mathbf e_{x'} = w{\mathbf e_x}, &&\mathbf e_{y'} = a {\mathbf e_y}, && \mathbf e_{z'} = a{\mathbf e_z}\label{A5}.
\end{align}

Because the physical vector is an invariant geometric object  ($v^i \mathbf e_i = v^{i'} \mathbf e_{i'}$), the stretching of the basis vectors dictates the inverse transformation of the vector components. Denoting the arrays of components as $\mathbf v$ and $\mathbf v'$, they transform alongside the spatial differential operators as:    
\begin{align}
    \mathbf v = \mathbb A \cdot \mathbf v', && \nabla = \mathbb A^{-1} \cdot \nabla'.
\end{align}
\subsection{Electromagnetic Tensor Transformation}
\label{a2}

While Equation \eqref{A5} shows the rigorous transformation of vector fields in the Expanding Box Model framework, the magnetic field $\mathbf B$ is a pseudo-vector. In the ideal MHD approximation, the plasma is treated as a perfect conductor ($\sigma \to \infty$). Consequently, the ideal Ohm's law dictates that the electric field must completely vanish in the rest frame of the plasma ($\mathbf{E}' = \mathbf{0}$) to prevent infinite currents. Because the electric field is zero in the co-moving frame, the electromagnetic Faraday tensor $F_{\mu\nu}$ is determined exclusively by the magnetic field components.

In contrast with the classical case, the magnetic field is an observer-dependent spatial projection of the electromagnetic tensor $F_{\alpha \beta}$. Thus, in a covariant framework, the fundamental coordinate transformation is applied to the rank-2 Faraday tensor $F_{\mu\nu}$. Consequently, the magnetic field must be recovered as an observer-dependent spatial projection of this tensor. The contravariant co-moving components of the magnetic field measured by a specific observer with four-velocity $u^{\mu'}$ are defined as ~\citep{Lichnerowicz1967,MTW}
\begin{equation}
    B^{\mu'} = \frac{1}{2\sqrt{-g}}\epsilon^{\mu'\nu'\alpha'\beta'}u_{\nu'} F_{\alpha'\beta'},
\end{equation}
where $\epsilon^{\mu'\nu'\alpha'\beta'}$ is the antisymmetric Levi-Civita symbol, and $g = -w^2 a^4$ is the metric determinant.

In the non-inertial co-moving frame, we explicitly evaluate the field as measured by an observer locally at rest with the expanding plasma parcel. This choice of observer dictates a four-velocity $u^{\mu'} = (1,0,0,0)$. Thus, the only non-zero term is $u_{t'} = -1$. As the only non-zero components are when $\nu'=t'$, the spatial components reduce to
\begin{equation}
    B^{i'} = \frac{1}{2 w a^2} \epsilon^{t'i'j'k'} F_{j'k'}.
\end{equation}
Expanding this expression for each spatial coordinate and utilizing the antisymmetry of the electromagnetic tensor ($F_{j'k'} = -F_{k'j'}$), we obtain:
\begin{align}
    B^{x'} = \frac{1}{w a^2} F_{y'z'}, \quad B^{y'} = \frac{1}{w a^2} F_{z'x'}, \quad B^{z'} = \frac{1}{w a^2} F_{x'y'}. \label{eq_b_comoving_F}
\end{align}
To relate these components to the inertial laboratory frame, we apply the covariant tensor transformation
\begin{equation}
    F_{\alpha'\beta'} = \frac{\partial x^\mu}{\partial x^{\alpha'}} \frac{\partial x^\nu}{\partial x^{\beta'}} F_{\mu \nu}.
\end{equation}
Using the EBM coordinate scalings given by Equation \eqref{A5}, we obtain $x = w x'$, $y = a y'$, $z = a z'$. The spatial components of the co-moving electromagnetic tensor are
\begin{align}
    F_{y'z'} = a^2 F_{yz}, \quad F_{z'x'} = a w F_{zx}, \quad F_{x'y'} = a w F_{xy}.
\end{align}
In the inertial frame, the classical definition applies directly ($B_x = F_{yz}$, $B_y = F_{zx}$, $B_z = F_{xy}$). Substituting these relations back into Equation \eqref{eq_b_comoving_F}, we obtain the non-inertial magnetic field as
\begin{align}
    B^{x'}   = \frac{B_x}{w}, && B^{y'} = \frac{B_y}{a},&&  B^{z'} =  \frac{B_z}{a}.
\end{align}
In vector form, we finally obtain the transformation of the magnetic field as
\begin{equation}
    \mathbf B = \mathbb A \cdot \mathbf B'.
\end{equation}

The transformation for the electric field is analogous, yielding $\mathbf E = \mathbb A \cdot \mathbf E'$.

\subsection{Relativistic EBM-MHD Equations}
\label{a3}
Using this geometric formalism, we derive the relativistic ideal MHD equations from the fundamental covariant conservation laws~\citep{Lichnerowicz1967,Anile1990}. These equations represent the spacetime generalizations of classical fluid conservation, and are given by
\begin{align}
    \nabla_{\mu'} J^{\mu'} &= 0, \label{cons_mass}\\
    \nabla_{\mu'} {}^*F^{\mu'\nu'} &= 0, \label{cons_faraday}\\
    \nabla_{\mu'} \left( T_\text{F}^{\mu'\nu'} + T_\text{EM}^{\mu'\nu'} \right) &= 0, \label{cons_energy}
\end{align}
where $\nabla_{\mu'}$ denotes the covariant derivative, which accounts for the inertial forces through the Christoffel symbols. Equation \eqref{cons_mass} represents the conservation of particles, where the mass four-current is $J^{\mu'} = \rho u^{\mu'}$, with $\rho$ being the proper mass density and $u^{\mu'}$ the plasma four-velocity. Equation \eqref{cons_faraday} corresponds to the homogeneous Maxwell equation, expressed through the dual electromagnetic tensor ${}^*F^{\mu'\nu'}$. Lastly, Equation \eqref{cons_energy} expresses the conservation of total energy and momentum. The stress-energy tensor in the MHD case is the sum of the fluid component $T_\text{F}^{\mu'\nu'}$ and the electromagnetic component $T_\text{EM}^{\mu'\nu'}$, defined respectively as
\begin{align}
    T_\text{F}^{\mu'\nu'} = (p + \rho) u^{\mu'} u^{\nu'} + p g^{\mu'\nu'},\\
    T_\text{EM}^{\mu'\nu'} = \frac{1}{4 \pi} \left (F^{\mu \lambda} F^\nu_{\lambda} - \frac{1}{4}g^{\mu \nu} F^{\alpha \beta} F_{\alpha \beta} \right ),
\end{align}
where $p$ is the isotropic fluid pressure and $F^{\mu\nu}$ is the standard electromagnetic tensor, and, due to the ideal MHD condition ($\mathbf{E}'=\mathbf{0}$), the electromagnetic stress-energy tensor is governed by the magnetic field contributions.

By evaluating the covariant derivatives defined in Equations~\eqref{cons_mass}-\eqref{cons_energy} using the Bianchi Type I metric given by Equation ~\eqref{metric}, the effects of solar wind expansion naturally arise within the framework of ideal MHD. 

For the conservation of mass, we expand the four-divergence using the fundamental identity for a vector $\nabla_{\mu'} J^{\mu'} = \frac{1}{\sqrt{-g}} \partial_{\mu'}(\sqrt{-g} J^{\mu'})$. Since the metric determinant $g = -w^2 a^4$ depends exclusively on the temporal coordinate $t'$, the spatial derivatives act only on the current. Defining the mass four-current as $J^{\mu'} = \rho u^{\mu'}$, where $\rho$ is the proper mass density and $u^{\mu'} = (\gamma, u^{i'})$ is the four-velocity decomposed into the Lorentz factor $\gamma$ and the spatial components $ u^{i'}$, we obtain the relativistic continuity equation in the co-moving frame, given by Equation \eqref{eq_cont_rel}:
\begin{equation}
    \gamma \partial_{t'}\rho + \rho \partial_{t'} \gamma + \partial_{i'} (\rho u^{i'}) = -\text{Tr}(\mathbb H) \gamma \rho, \label{eq_cont_rel}
\end{equation}
where
    $\mathbb H = \text{diag}\left (\frac{\dot w}{w}, \frac{ \dot a}{a}, \frac{\dot a}{a} \right )$
is the expanding rate tensor, in analogy to \citet{Nariyuki2015}, and $\gamma$ is the Lorentz factor. Here, the geometric dilution of plasma density arises naturally from the temporal evolution of the metric. Similarly, we expand the covariant divergence of the anti-symmetric dual electromagnetic tensor to obtain Faraday's Law in the expanding frame. Separating the temporal ($t'$) and spatial ($i', j' \in \{x', y',z'\}$) components and defining $B^{t'}$ and $B^{i'}$ as the temporal and spatial components of the magnetic field in the co-moving frame, we obtain the the relativistic, expanding Faraday's Law, given by Equation \eqref{eq_faraday_rel}:
\begin{equation}
    \begin{aligned}[b]
        \partial_{t'} (\gamma B^{i'} - u^{i'} B^{t'})+  \partial_{j'} (u^{j'} B^{i'} - u^{i'}B^{j'}) \\=  - (\gamma B^{i'} - u^{i'} B^{t'}) \text{Tr}(\mathbb H),\label{eq_faraday_rel}
    \end{aligned}
\end{equation}

The evolution of the fluid momentum is described by the spatial components of the energy-momentum conservation, $\nabla_{\mu'} T^{\mu' i'} = 0$. In differential geometry, the covariant derivative of a rank-2 tensor introduces terms proportional to the Christoffel symbols, $\Gamma^{i'}_{\alpha'\beta'} T^{\alpha'\beta'}$. These terms represent the inertial forces acting on the plasma due to spacetime stretching. Defining the relativistic total enthalpy $W = \rho(1+\epsilon) + p + B^2/4\pi$ and total pressure $P_T = p + B^2/8\pi$, the fully relativistic momentum equation is given by Equation \eqref{eq_mom_rel}:
\begin{equation}
\begin{aligned}[b]
    \partial_{t'} \left( W \gamma^2 v^{i'} - \frac{B^{t'} B^{i'}}{4\pi} \right)\\+ \partial_{j'} \left( W \gamma^2 v^{i'}v^{j'} + P_T g^{i'j'} - \frac{B^{i'} B^{j'}}{4\pi} \right)\\ 
    = - \left( \text{Tr}(\mathbb{H}) \delta_{k'}^{i'} + 2\mathbb{H}_{k'}^{i'} \right) \left[ W \gamma^2 v^{k'} - \frac{B^{t'} B^{k'}}{4\pi} \right], \label{eq_mom_rel}
\end{aligned}
\end{equation}
where $\delta_{k'}^{i'}$ is the Kronecker delta. Finally, we project the covariant derivative of the total energy-momentum tensor along the four-velocity, $u_{\nu'} \nabla_{\mu'} T^{\mu'\nu'} = 0$, where the electromagnetic field does no work in the ideal MHD rest frame ($F^{\mu'\nu'}u_{\nu'} = 0$). Assuming a polytropic ideal gas equation of state $p = (\Gamma - 1)\rho \epsilon$, where $\epsilon$ is the specific internal energy and $\Gamma$ the adiabatic index, we obtain the covariant adiabatic pressure equation given by Equation \eqref{eq_press_rel}:
\begin{equation}
    \gamma \partial_{t'}p + \gamma v^{i'}\partial_{i'}p + \Gamma p \partial_{t'} \gamma + \gamma \Gamma p \partial_{i'}v^{i'} = - \Gamma p \gamma \text{Tr}(\mathbb H). \label{eq_press_rel}
\end{equation}

Equations \eqref{eq_cont_rel}-\eqref{eq_press_rel} represent the complete set of ideal MHD equations for a relativistically expanding plasma. While macroscopic relativistic velocities are generally not required for solar wind applications, deriving the system from a covariant foundation ensures that the geometric coupling between the vector fields and the expanding spacetime is mathematically consistent. Beyond the heliospheric scenario, this framework could open up new possibilities for exploring broader expanding astrophysical phenomena in relativistic regimes.

\section{Transformation of the EBM-MHD Equations to the Inertial Frame}
\label{appendixB}

Here, we detail the algebraic transformations required to map the non-relativistic EBM-MHD equations from the co-moving frame $S'$ to the inertial laboratory frame $S$, since these frames are not equivalent in our framework. The fundamental transformations connecting the frames, as dictated by the stretching basis vectors, are $\mathbf{B}' = \mathbb{A}^{-1} \cdot \mathbf{B}$, $\mathbf{v}' = \mathbb{A}^{-1} \cdot \mathbf{v}$, and $\nabla' = \mathbb{A} \cdot \nabla$. We also write $\dot{\mathbb{A}}^{-1} = - \mathbb{A}^{-1} \cdot \mathbb{H}$.

\subsection{Momentum Equation}
Starting from the co-moving momentum equation \eqref{eq_mom_nr}, the convective derivative term transforms as:
\begin{equation}
\begin{split}
    \rho \left [ \frac{\partial \mathbf{v}'}{\partial t} + (\mathbf{v}' \cdot \nabla) \mathbf{v}' \right ] \\ = \rho \left [\frac{\partial }{\partial t}(\mathbb{A}^{-1} \cdot \mathbf{v}) + [(\mathbb{A}^{-1} \cdot \mathbf{v}) \cdot (\mathbb{A} \cdot \nabla)](\mathbb{A}^{-1} \cdot \mathbf{v}) \right ]\\ = \rho \mathbb{A}^{-1}\cdot \left ( \frac{\partial \mathbf{v}}{\partial t} - \mathbb{H} \cdot \mathbf{v} + (\mathbf{v} \cdot \nabla) \mathbf{v}\right ). \label{app_mom_conv}
\end{split}
\end{equation}
The gradient and magnetic tension terms transform respectively as:
\begin{align}
    g^{i'j'} \nabla' p &=  \mathbb{A}^{-2} \cdot (\mathbb{A} \cdot \nabla) p =  \mathbb{A}^{-1} \cdot \nabla p,\\
    g^{i'j'} \nabla' B'^2 &= \mathbb{A}^{-2} \cdot (\mathbb{A} \cdot \nabla) B^2 = \mathbb{A}^{-1}\cdot \nabla B^2, \\
    (\mathbf{B}' \cdot \nabla')\mathbf{B}' &= \left [\mathbf B \cdot \nabla \right ](\mathbb{A}^{-1} \cdot \mathbf{B}) = \mathbb{A}^{-1} \cdot (\mathbf{B} \cdot \nabla) \mathbf{B},\\
        2 \rho \mathbb{H} \cdot \mathbf{v}' & = 2 \rho \mathbb{H} \cdot (\mathbb{A}^{-1} \cdot \mathbf{v}) = \mathbb{A}^{-1} \cdot (2 \rho \mathbb{H} \cdot \mathbf{v}).
\end{align}
where we have used $B'^2 = B^2$ as the magnitude of a vector remains the same, and $g^{i'j'} = \mathbb{A}^{-2}$. 
Multiplying the entire equation by $\mathbb{A}$ from the left and simplifying, we obtain
\begin{equation}
\begin{split}
        \rho \frac{\partial \mathbf{v}}{\partial t} + \rho (\mathbf{v} \cdot \nabla) \mathbf{v} + \nabla p + \frac{1}{8 \pi}\nabla B^2\\ - \frac{1}{4 \pi}(\mathbf{B} \cdot \nabla) \mathbf{B} = - \rho \mathbb{H} \cdot \mathbf{v}, 
\end{split}
\end{equation}

\subsection{Faraday's Law}
Transforming the expanded co-moving induction equation, we obtain:
\begin{equation}
\begin{split}
        \mathbb{A}^{-1} \cdot \left [\frac{\partial \mathbf{B}}{\partial t} - \mathbb{H} \cdot \mathbf{B} + (\nabla \cdot \mathbf{v})\mathbf{B} + (\mathbf{v} \cdot \nabla) \mathbf{B} - (\mathbf{B} \cdot \nabla) \mathbf{v} \right ]  \\= - \text{Tr}(\mathbb{H})  \mathbb{A}^{-1} \cdot \mathbf{B}.
\end{split}
\end{equation}
Multiplying by $\mathbb{A}$ and grouping the terms containing $\mathbb{H}$, we derive the inertial geometric source term:
\begin{equation}
    \frac{\partial \mathbf{B}}{\partial t} + (\nabla \cdot \mathbf{v})\mathbf{B} + (\mathbf{v} \cdot \nabla) \mathbf{B} - (\mathbf{B} \cdot \nabla) \mathbf{v} = - \left [\frac{\dot a}{a} \mathbb{L} + \frac{\dot w}{w} \mathbb{T}\right ]\cdot \mathbf{B}.
\end{equation}

\section{Solution for time-dependent trajectory $R(t)$}
\label{appendixC}

Here, we provide the mathematical framework detailing the mathematical equivalence between the temporal evolution of a plasma parcel and its spatial parameterization in the EBM. To find the explicit form $R(t)$ in the velocity profile given by Equation \eqref{velprofile}, we must integrate the time $dt' = dR'/V(R)$ to an arbitrary time $t$. Defining $\Delta V = V_\infty - V_0$ and solving the integral, the transit time as a function of distance, $t(R)$, is obtained as
\begin{equation}
    t(R) = \frac{R - R_0}{V_\infty} + \frac{\Delta V R_0}{V_\infty^2} \ln  \left (\frac{V_\infty R - \Delta V R_0 }{V_0 R_0} \right ).
\end{equation}

To obtain the trajectory $R(t)$, this equation must be inverted. We begin by multiplying the entire expression by $V_\infty^2 /(\Delta V R_0)$ and defining the characteristic expansion rate $\tau = V_\infty^2 / (\Delta V R_0)$ and the velocity ratio $\xi = V_0 / \Delta V$. Defining $y = (V_\infty R - \Delta V R_0) / (V_0 R_0)$, the equation simplfies significantly
\begin{equation}
    \tau t = (y\xi + 1) - (\xi + 1) + \ln(y) = \xi(y - 1) + \ln(y).
\end{equation}
Rearranging the terms, exponentiating both sides, and multiplying by $\xi$, the equation is written as
\begin{equation}
    \xi e^{\tau t + \xi} = (\xi y) e^{\xi y}.
\end{equation}
By definition, the inverse function is the Lambert W Function. Let $x(t) = \xi \exp (\tau t + \xi)$. The equation becomes $\xi y = W (x(t))$, allowing us to solve $y = \frac{1}{\xi} W(x(t))$. Restoring the initial variables, we arrive at the exact trajectory $R(t)$ as 
\begin{equation}
    R(t) = R_0 \frac{\Delta V}{V_\infty} [1 + W(z(t))].
\end{equation}
To show that our parameterization in terms of $a = R/R_0$ incurs no loss of generality regarding the non-constant acceleration profile, we compute $\gamma_a = \frac{\dot R(t)}{R_0}$ and $\gamma_w = \frac{dV}{dR}$. Using the derivative property of the Lambert W function, $\frac{d}{dx}W(x) = W(x)/[x(1+W(x))]$, and applying the chain rule $\frac{dx}{dt} = \tau x(t)$, $\dot R(t)$ is obtained as
\begin{equation}
    \dot R(t) = V_\infty \frac{W(x(t))}{1 + W(x(t))}.
\end{equation}
Using the definition of $\gamma_a$, we obtain
\begin{equation}
    \gamma_a(t) = \frac{V_\infty^2}{R_0 \Delta V} \frac{W(x(t))}{[1 + W(x(t))]^2}.
\end{equation}

On the other hand, evaluating the spatial parametrization $\gamma_a(a) = V(a)/(a R_0)$ and substituting $a(t) = R(t)/R_0$, we obtain
\begin{equation}
    \gamma_a(a)  = \frac{V_\infty^2}{R_0 \Delta V} \frac{W(x(t))}{[1 + W(x(t))]^2},
\end{equation}
thus showing that mapping from time dependencies to spatial dependencies does not require additional approximations. An analogous procedure shows the same equivalence for the acceleration rate $\gamma_w$.

\section{Comparison with other Expanding Box Models in the literature}
\label{appendixD}

To demonstrate the consistency and generalization provided by our covariant framework, we compare our derived MHD equations with the standard Expanding Box Model (EBM) formulations widely used in the literature. As many authors have derived the MHD set of equations in the EBM frame using different approaches, this section helps clarify some discrepancies in the notation and approaches. We will focus our comparison on the magnetic field and momentum equations, as scalar fields are equivalent across reference frames.

\subsection{Comparison with Grappin et al. (1993)}

We first compare our MHD equations with the standard Expanding Box Model formulation originally derived by \citet{Grappin1993}. The classical EBM assumes a plasma parcel moving at a constant radial speed $V_0$, with radial acceleration vanishing ($w = 1 \implies \gamma_w = 0$), and thus expansion is purely in the transverse directions. 

In their derivation, \citet{Grappin1993} apply a coordinate transformation to a comoving frame, as their gradient operator is$\nabla = [\partial/\partial\tilde{x}, (1/a)\partial/\partial\tilde{y}, (1/a)\partial/\partial\tilde{z}]$. Their vector fields ($\mathbf{v}$ and $\mathbf{B}$) are expressed in terms of an inertial reference frame. Their resulting momentum and magnetic field equations are
\begin{equation}
\begin{aligned}
\frac{\partial \mathbf v}{\partial t} + (\mathbf v \cdot \nabla)\mathbf v -  \frac{1}{\rho}(\mathbf B \cdot \nabla)\mathbf B \\+ \frac{1}{\rho}\nabla \left (P + \frac{B^2}{2} \right ) = - \frac{\dot a}{a} \mathbb T \cdot \mathbf u,
\end{aligned}
\end{equation}

\begin{equation}
\begin{aligned}
\frac{\partial \mathbf B}{\partial t} + (\mathbf v \cdot \nabla) \mathbf B - (\mathbf B \cdot \nabla) \mathbf v  + \mathbf B (\nabla \cdot \mathbf v) = - \frac{\dot a}{a} \mathbb L \cdot \mathbf B,
\end{aligned}
\end{equation}
where $\mathbb{T} = \text{diag}(0, 1, 1)$ and $\mathbb{L} = \text{diag}(2, 1, 1)$. 

In contrast, our covariant formulation is constructed in the non-inertial, co-moving frame. By applying the constant velocity limit ($\gamma_w \to 0$) to our model, the expansion rate tensor $\mathbb H$ is simplified to $\mathbb H = \frac{\dot a}{a} \mathbb T$. We recover the equations deduced by \citet{Grappin1993} when using this transformation in Equations \eqref{eq_mom_s} and \eqref{eq_faraday_s}. As a result, despite being formulated in the non-inertial comoving frame, our covariant approach recovers the physics of the classical inertial EBM. Our derivation also unifies the $\mathbb{T}$ and $\mathbb{L}$ matrices into a single expansion tensor $\mathbb{H}$.

\subsection{Comparison with Tenerani \& Velli 2013}

\citet{Tenerani2013} extended the classical EBM by incorporating a longitudinal acceleration profile, parameterized by the rate $\dot{w}/w$. To capture the coupling between wave dynamics and the heliospheric structure, their formulation retains the spatial gradients of the macroscopic background. As a consequence, their equations for velocity fluctuations $\mathbf u$, the transverse magnetic field $\mathbf B_\perp$, and density $\rho$ took the following form (their Equations 1, 3, and 4):
\begin{equation}
\begin{aligned}
\frac{\partial \mathbf{u}}{\partial t} + \mathbf{u} \cdot \nabla \mathbf{u} = -\frac{1}{\rho}\left[ \nabla \left(p + \frac{B^2}{2}\right) - \mathbf{B} \cdot \nabla \mathbf{B} \right] - \hat{P} \cdot \mathbf{u} \\ + \frac{1}{\rho_0}\frac{B_0 \mathbf{B}_\perp}{R} + \left[ \hat{\mathbf{x}} \frac{[\rho_0(T-T_0) + T_0(\rho-\rho_0)]}{\rho_0} \right] \frac{\Gamma-1}{2U} \\ - \frac{\mathbf{B}_\perp B_0}{\rho_0}\frac{1}{2V(R)}\left(2\frac{\dot{a}}{a} + \frac{\dot{w}}{w}\right), 
\end{aligned}
\end{equation}

\begin{equation}
\begin{aligned}
\frac{\partial \mathbf{B}_\perp}{\partial t} &= [\nabla \times (\mathbf{u} \times \mathbf{B})]_\perp - \hat{B} \cdot \mathbf{B}_\perp - \frac{\mathbf{u}_\perp B_0}{R},
\end{aligned}
\end{equation}

\begin{equation}
    \frac{\partial \rho}{\partial t} + \nabla \cdot \rho \mathbf u = - \rho \left (2 \frac{\dot a}{a} + \frac{\dot w}{w}  \right ) + u_x \frac{\rho_0}{V(R)} \frac{\dot w}{w}.
\end{equation}
This formulation identifies the total expansion rate, written in our formulation as $\text{Tr}(\mathbb H)$. On the other hand, the retention of spatial background gradients introduces explicit curvature effects ($\propto R^{-1}$) and convective terms ($\propto u_x$). These terms break the spatial symmetry required in standard EBM frameworks. 

In contrast, our formulation defines a limit  $(L \ll R)$ through the expanding metric where inhomogeneities are neglected. In this limit, curvature effects are omitted, and expansion effects are represented purely in the non-inertial forces. Consequently, our momentum and magnetic field equations isolate all expansion effects into the $\mathbb H$ effects, while geometric wave reflection is purely driven by the coupling matrix $\mathbb M$ instead of plasma gradients. Thus, our covariant model successfully captures the inertial expansion effects while ensuring strict local homogeneity.

\subsection{Comparison with Del Zanna et al. (2015)}

To implement the Expanding Box Model into conservative numerical simulations, \citet{Del_Zanna2015} formulated the ideal EBM-MHD equations utilizing a covariant, general relativistic framework. By treating the expanding comoving frame as a time-dependent spacetime, they defined the macroscopic equations in terms of the metric tensor $g_{ij}$ and its determinant $\sqrt{g}$.

In their formulation, they consider the classical EBM regime with constant velocity $V_0$). Thus, they define their metric tensor to capture purely transverse expansion, where  $g_{ij} = \text{diag}\{1, a^2, a^2\}$. From here, the MHD system is written in a generalized conservative form $\partial_t \mathcal{U} + \partial_i \mathcal{F}^i = \mathcal{S}$. The state vector $\mathcal{U}$, the fluxes $\mathcal{F}^i$, and the source terms $\mathcal{S}$ are given by
\begin{equation}
    \mathcal{U} = \sqrt{g} \begin{bmatrix} \rho \\ \rho v_j \\ E_t \\ B^j \end{bmatrix}, \quad \mathcal{F}^i = \sqrt{g} \begin{bmatrix} \rho v^i \\ \rho v^i v_j - B^i B_j + p_t \delta^i_j \\ (E_t + p_t)v^i - (v_k B^k)B^i \\ v^i B^j - v^j B^i \end{bmatrix},
\end{equation}
\begin{equation}
    \mathcal{S} = \sqrt{g} \begin{bmatrix} 0 \\ \frac{1}{2}(\rho v^i v^k - B^i B^k + p_t g^{ik})\partial_j g_{ik} \\ -\frac{1}{2}(\rho v^i v^j - B^i B^j + p_t g^{ij})\partial_t g_{ij} \\ 0 \end{bmatrix}.
\end{equation}
By expanding these equations into orthonormal Cartesian components ( $v^2 = v_y/a$ and $B^2 = B_y/a$), this system is equivalent to our equations \eqref{eq_cont_s}-\eqref{eq_faraday_s} evaluated in the constant velocity limit. However, because their metric $g_{ij}$ scales only the transverse plane, their covariant model assumes zero longitudinal acceleration. Our model generalizes this by extending the covariant foundation established by \citet{Del_Zanna2015}, extending the tensor $g_{ij}$ to capture longitudinal acceleration. Thus, our formulation extends their geometrically this formalism into the solar wind acceleration region, successfully capturing non-WKB reflection ($\mathbb{M}_\perp$) while ensuring that all non-inertial expansion effects remain confined to the expansion tensor $\mathbb{H}$.

\subsection{Comparison with Nariyuki 2015}

Another significant effort to incorporate longitudinal acceleration into the Expanding Box Model was presented by \citet{Nariyuki2015}. Through a coordinate transformation parameterized by a time-dependent scale matrix $\mathbf{W} = \text{diag}(w, a, a)$, they deduced a set of modified MHD equations to describe an accelerating plasma parcel.

This formulation identifies the expansion tensor $\mathbb H$, written in their work as $\mathbf D$. Based on this, their equations for momentum and magnetic field are
\begin{equation}
\begin{aligned}
\rho \frac{\partial \mathbf{u}'}{\partial t} + \rho \mathbf{u}' \cdot \nabla' \mathbf{u}' + \nabla' \left( p + \frac{|\mathbf{b}|^2}{2\mu_0} \right) \\- \frac{1}{\mu_0}(\mathbf{b} \cdot \nabla')\mathbf{b} = -\mathbf{D} \rho \mathbf{u}' - \ddot{\mathbf{W}} \rho \mathbf{x}',
\end{aligned}
\end{equation}

\begin{equation}
\begin{aligned}
        \frac{\partial \mathbf{b}}{\partial t} - \nabla' \times (\mathbf{u}' \times \mathbf{b}) + \nabla' \times \left( \frac{1}{\rho \mu_0} (\nabla' \times \mathbf{b}) \times \mathbf{b} \right) \\= -((\text{tr}\mathbf{D})\mathbf{I} - \mathbf{D})\mathbf{b},
\end{aligned}
\end{equation}

where $\nabla' = \left (\frac{1}{w} \frac{\partial }{\partial x'},\frac{1}{a} \frac{\partial }{\partial y'},\frac{1}{a} \frac{\partial }{\partial z'}  \right )$ is their transformed spatial gradient.

In the system presented by \citet{Nariyuki2015}, their magnetic field equation is equivalent to the magnetic field equation derived in our model for the inertial frame (Equation \eqref{eq_faraday_s}). 

However, there is a difference in how the vector fields are mathematically treated. In their formulation, the spatial coordinates and the velocity field are explicitly transformed to the comoving frame ($\mathbf{x}'$ and $\mathbf{u}'$), whereas the magnetic field $\mathbf{b}$ remains unprimed. This indicates that $\mathbf{b}$ is treated as a standard Galilean vector evaluated in the co-moving frame. As a consequence, in their momentum equation, the magnetic field terms are forced to couple the unprimed Galilean components with the transformed gradient $\nabla'$.

In contrast, our covariant formulation eliminates this asymmetry. Here, all physical quantities are transformed as proper geometric tensors. Here, the momentum equation maintains its structure without exposing the scale factors in the spatial operators. The term $-\ddot{\mathbf{W}} \rho \mathbf{x}'$ is similarly absorbed by the comoving frame. Thus, while perfectly recovering the physical decay rates identified by \citet{Nariyuki2015}, our model confines the non-inertial effects strictly to the expansion tensor $\mathbb{H} \cdot \mathbf{u}$.

\subsection{Comparison with Echeverría-Veas et al. 2023}

Recently, \citet{Echeverria-Veas2023} deduced the ideal MHD system of equations in the EBM framework by performing the coordinate transformations on an expanding Vlasov equation.

In their constant-velocity formulation, the momentum and magnetic field equations are expressed as
\begin{equation}
\begin{aligned}
\frac{\partial \mathbf{u}}{\partial t} + (\mathbf{u} \cdot \nabla)\mathbf{u} + \frac{1}{\rho}\nabla p + \frac{1}{8\pi\rho}[(\mathbb{A}^{-1} \cdot \nabla)]B^2 \\
- \frac{1}{4\pi\rho}[\mathbf{B} \cdot (\mathbb{A}^{-1} \cdot \nabla)]\mathbf{B} = -\frac{V_0}{aR_0}\mathbb{T} \cdot \mathbf{u},
\end{aligned}
\end{equation}

\begin{equation}
\begin{aligned}
\frac{\partial \mathbf{B}}{\partial t} + (\nabla \cdot \mathbf{u})\mathbf{B} + (\mathbf{u} \cdot \nabla)\mathbf{B} \\
- [\mathbf{B} \cdot (\mathbb{A}^{-1} \cdot \nabla)](\mathbb{A} \cdot \mathbf{u}) = -\frac{V_0}{aR_0}\mathbb{L} \cdot \mathbf{B},
\end{aligned}
\end{equation}
where $\frac{V_0}{aR_0} = \frac{\dot a}{a}$ in the constant-velocity limit, and $\nabla$ is their co-moving spatial gradient. By evaluating our model in the constant-velocity limit, our expansion tensor reduces to $\mathbb{H} = \frac{\dot a}{a}\mathbb{T}$, recovering the velocity and magnetic field damping. However, their coordinate transformation relies on applying the inverse matrix $\mathbb{A}^{-1}$ directly to the co-moving spatial gradient $\nabla$. Because the magnetic field is not transformed as a proper tensor, the transformation matrix explicitly appears in the magnetic pressure gradient, the magnetic tension, and the convective induction term $[\mathbf{B} \cdot (\mathbb{A}^{-1} \cdot \nabla)](\mathbb{A} \cdot \mathbf{u})$, losing their symmetric form. By properly transforming the magnetic field in our covariant formulation, our framework resolves this algebraic complication.

\bibliography{references}{}
\bibliographystyle{aasjournalv7}

\end{document}